\pdfoutput=1
\documentclass{jfm1}
\usepackage{natbib}

\newcommand{\enma}[1]   {\ensuremath{#1}}

\newcommand{\non}{\nonumber}

\newcommand{\beq}{\begin{equation}}
\newcommand{\eeq}{\end{equation}}
\newcommand{\bseq}{\begin{subequations}}
\newcommand{\eseq}{\end{subequations}}
\newcommand{\beqn}{\begin{eqnarray}}
\newcommand{\eeqn}{\end{eqnarray}}
\newcommand{\ba}{\begin{array}}
\newcommand{\ea}{\end{array}}
\newcommand{\bct}{\begin{center}}
\newcommand{\ect}{\end{center}}
\newcommand{\btmz}{\begin{itemize}}
\newcommand{\etmz}{\end{itemize}}
\newcommand{\benum}{\begin{enumerate}}
\newcommand{\eenum}{\end{enumerate}}

\newcommand{\trace}     {\enma{\mathrm{trace}}}

\newcommand{\bv}{{\bf v}}

\newcommand{\matbegin}{
        \left[
}
\newcommand{\matend}{
        \right]
}

\newcommand{\tbo}[2]{
  \matbegin \begin{array}{c}
       #1 \\ #2
       \end{array} \matend }

\newcommand{\obt}[2]{
  \matbegin \begin{array}{cc}
       #1 & #2
       \end{array} \matend }

\newcommand{\tbt}[4]{
  \matbegin \begin{array}{cc}
       #1 & #2 \\ #3 & #4
       \end{array} \matend }

\newcommand{\tbth}[6]{
  \matbegin \begin{array}{ccc}
       #1 & #2 & #3\\ #4 & #5 & #6
       \end{array} \matend }

\newcommand{\be}{\begin{equation}}
\newcommand{\ee}{\end{equation}}

\newcommand{\cplxs}{ C\kern -.35em \rule{0.03 em}{.7 ex}~   }

\def\complex{\hbox{C\kern -.45em \rule{0.03 em}{1.5 ex}}~}

\newcommand{\bi}{\begin{itemize}}
\newcommand{\ei}{\end{itemize}}

\usepackage{epsfig,amssymb,graphics,color,latexsym,subfigure,psfrag,amsmath,pifont,mathtools}

\mathtoolsset{showmanualtags}

\ifCUPmtlplainloaded \else
  \checkfont{eurm10}
  \iffontfound
    \IfFileExists{upmath.sty}
      {\typeout{^^JFound AMS Euler Roman fonts on the system,
                   using the 'upmath' package.^^J}%
       \usepackage{upmath}}
      {\typeout{^^JFound AMS Euler Roman fonts on the system, but you
                   dont seem to have the}%
       \typeout{'upmath' package installed. JFM.cls can take advantage
                 of these fonts,^^Jif you use 'upmath' package.^^J}%
      }
  \else
  \fi
\fi

\ifCUPmtlplainloaded \else
  \checkfont{msam10}
  \iffontfound
    \IfFileExists{amssymb.sty}
      {\typeout{^^JFound AMS Symbol fonts on the system, using the
                'amssymb' package.^^J}%
       \usepackage{amssymb}%
       \let\le=\leqslant  
         
      }{}
  \fi
\fi

\ifCUPmtlplainloaded \else
  \IfFileExists{amsbsy.sty}
    {\typeout{^^JFound the 'amsbsy' package on the system, using it.^^J}%
     \usepackage{amsbsy}}
    {}
\fi

\newcommand{\bA}{\mathbf{A}}
\newcommand{\bB}{\mathbf{B}}
\newcommand{\bC}{\mathbf{C}}
\newcommand{\bF}{\mathbf{F}}

\newcommand{\bH}{\mathbf{H}}
\newcommand{\bL}{\mathbf{L}}
\newcommand{\bM}{\mathbf{M}}
\newcommand{\bN}{\mathbf{N}}
\newcommand{\bP}{\mathbf{P}}

\newcommand{\bS}{\mathbf{S}}
\newcommand{\bT}{\mathbf{T}}
\newcommand{\bX}{\mathbf{X}}
\newcommand{\bY}{\mathbf{Y}}
\newcommand{\bZ}{\mathbf{Z}}
\newcommand{\bI}{\mathbf{I}}
\newcommand{\ds}{\displaystyle}
\newcommand{\bphi}{\mbox{\boldmath$\phi$}}
\newcommand{\bvphi}{\mbox{\boldmath$\varphi$}}
\newcommand{\bpsi}{\mbox{\boldmath$\psi$}}

\newcommand{\btau}{\mbox{\boldmath$\tau$}}

\newcommand{\bDelta}{\mbox{\boldmath$\Delta$}}

\newcommand{\bd}{{\bf d}}

\newcommand{\mrd}{\mathrm{d}}

\newcommand{\mri}{\mathrm{i}}

\newcommand{\pt}{\partial_t}
\newcommand{\py}{\partial_y}
\newcommand{\pyy}{\partial_{yy}}
\newcommand{\We}{W\!e}
\newcommand{\bPi}{\bar{\Pi}}
\newcommand{\inprod}[2]{\left< #1 , #2 \right>}

\title[Frequency responses in streamwise-constant channel flows of Oldroyd-B fluids]
{Frequency responses of streamwise-constant
    \\
perturbations in channel flows of Oldroyd-B
    \\
    fluids}

\author[N.\ Hoda, M.\ R.\ Jovanovi\'c, and S.\ Kumar]%
{N\ls A\ls Z\ls I\ls S\ls H \ls \ls H\ls O\ls D\ls A$^1$,
M\ls I\ls H\ls A\ls I\ls L\ls O \ls \ls R.\ls \ls J\ls O\ls V\ls A\ls N\ls O\ls V\ls I\ls \'C$^2$,
\and
S\ls A\ls T\ls I\ls S\ls H \ls \ls K\ls U\ls M\ls A\ls R\ls$^1$}
\affiliation{$^1$Department of Chemical Engineering and Materials Science, University of Minnesota,
Minneapolis, MN 55455, USA\\[\affilskip]
$^2$Department of Electrical and Computer Engineering, University of
Minnesota, \\ Minneapolis, MN 55455, USA}

\date{?? and in revised form ??}
\begin{document}

\maketitle

   \begin{abstract}
Non-modal amplification of disturbances in streamwise-constant
channel flows of Oldroyd-B fluids is studied from an input-output
point of view by analyzing the responses of the velocity components
to spatio-temporal body forces. These inputs into the governing
equations are assumed to be harmonic in the spanwise direction and
stochastic in the wall-normal direction and in time. An explicit
Reynolds number scaling of frequency responses from different
forcing to different velocity components is developed, showing the
same $Re$-dependence as in Newtonian fluids. It is found that some
of the frequency response components peak at non-zero temporal
frequencies. This is in contrast to Newtonian fluids, where
peaks are always observed at zero frequency, suggesting that
viscoelastic effects introduce additional timescales and promote
development of flow patterns with smaller time constants than in
Newtonian fluids. The temporal frequencies, corresponding to the
peaks in the components of frequency response, decrease with an
increase in viscosity ratio (ratio of solvent viscosity to total
viscosity) and show maxima for non-zero elasticity
number. Our analysis of the Reynolds-Orr equation demonstrates that
the energy-exchange term involving the streamwise/wall-normal
polymer stress component $\tau_{xy}$ and the wall-normal gradient of the streamwise velocity $\partial_y u$ becomes increasingly important relative to the
Reynolds stress term as the elasticity number increases, and is thus
the main driving force for amplification in flows with strong
viscoelastic effects.
   \end{abstract}

\section{Introduction}
    \label{sec.intro}

Complex dynamical responses arise in numerous viscoelastic fluid
flows~\cite[][]{Larson1992,Shaqfeh1996} and their study is important
from both fundamental and technological standpoints. From the former
standpoint, the inception and evolution of amplification of
disturbances in various flows involving viscoelastic fluids is not
well understood. Viscoelastic effects not only modify features
already present in Newtonian fluids, but also give rise to
completely new behavioral
patterns~\cite[][]{Larson2000,Groisman2000,Morozov2003a}. From the
latter standpoint, the study of dynamics in flows involving
polymeric fluids is of immense importance for polymer processing and
rheometry~\cite[][]{Bird1987,Larson1999}. Classical linear
hydrodynamic stability analysis is found to give misleading results
even for simple Couette and Poiseuille flows of Newtonian
fluids~\cite[][]{Schmid2007}. This failure of the classical
stability analysis is attributed to the non-normal nature of the
generators in the linearized governing
equations~\cite[][]{Trefethen1993,Grossmann2000,schhen01}. Linear
dynamical systems with non-normal generators can have solutions that
grow substantially at short times, even though they decay at long
times~\cite[][]{Gustavsson1991,Butler1992,redhen93}. Furthermore,
the non-normal nature of the underlying equations can lead to
significant amplification of ambient
disturbances~\cite[][]{Farrell1993,Bamieh2001,Mihailo2005} and
substantial decrease of stability
margins~\cite[][]{Trefethen1993,treemb05}. On some occasions, the
transient growth and amplification, which are overlooked in standard
linear stability analysis, could put the system in a regime where
nonlinear interactions are no longer negligible. These phenomena are
also expected to be important in Couette and Poiseuille flows of
viscoelastic fluids. In this paper, we investigate amplification of
disturbances in channel flows of Oldroyd-B fluids by performing a
frequency-response analysis.

Novel ways of describing fluid stability that allow quantitative
description of short-time behavior and disturbance amplification,
referred to as non-modal stability analysis, have emerged in the
last decade~\cite[][]{Schmid2007}. One approach is to study the
responses of the linearized Navier-Stokes equations (LNSE) to
external disturbances~\cite[][]{Farrell1993,Bamieh2001}.
\citet{Mihailo2005} have used this approach to study the effects of
external disturbances, in the form of body forces, on channel flows
of Newtonian fluids. Explicit Reynolds number dependence of the
components of the frequency response was derived. Based on this
finding, it was concluded that at higher Reynolds numbers,
wall-normal and spanwise disturbances have the strongest influence
on the flow field and the impact of these forces is largest on the
streamwise velocity. It was also found that the frequency responses
peak at different locations in the ($k_x,k_z$)-plane, where $k_x$
and $k_z$ are the streamwise and spanwise wavenumbers, indicating
the possibility of distinct amplification mechanisms.
We note that -- even in high-Reynolds-number regimes -- it is valid to examine the linearized equations to determine the fate of small-amplitude perturbations to the underlying base flow.

Recently, the present authors have extended the work
of~\citet{Mihailo2005} to viscoelastic
fluids~\cite*[][]{Nazish2008a}. Prior studies on transient growth
phenomena in viscoelastic fluids were reviewed there, and hence will
not be reviewed here for brevity. The aggregate effect of stochastic
disturbances in all the three spatial directions to all the three
velocity components, referred to as the ensemble-average energy
density, was investigated. It was found that the energy density
increases with an increase in elasticity number and a decrease in
viscosity ratio (ratio of solvent viscosity to total viscosity). In
most of the cases, streamwise-constant or nearly streamwise-constant
perturbations are most amplified and the location of maximum energy
density shifts to higher spanwise wavenumbers with an increase in
elasticity number and a decrease in viscosity ratio. However, prior
work on Newtonian fluids by~\citet{Mihailo2005} suggests that a
plethora of additional insight can be uncovered by analyzing the
componentwise spatio-temporal frequency responses. The componentwise
responses give information about the relative importance of the
three disturbances on the three velocity components. By analyzing
these frequency responses, the disturbance frequency corresponding
to maximum amplification can also be obtained.

In this paper, we examine the componentwise frequency responses in
streamwise-constant channel flows of Oldroyd-B fluids. This study
supplements a previous study by the authors~\citep{Nazish2008a},
where the aggregate effect of disturbances was examined, and helps
in understanding the relative importance of the disturbances on the
different velocity components.  Because the previous study focused
on aggregate effects, as parameterized by the energy density, it
leaves open the question of exactly which velocity components
are most amplified and which forcing components are responsible for this
amplification.  Furthermore, since the energy density is a time-integrated
quantity, it does not yield information about most amplified temporal
frequencies.  The present work addresses these issues, provides some explicit scaling relationships, and further investigates physical mechanisms.

Streamwise-constant three-dimensional perturbations are considered in this work as they are most amplified by the linearized dynamics. We derive an explicit Reynolds number scaling for the components of the frequency response. As in Newtonian
fluids, at higher Reynolds numbers the forces in the wall-normal and
spanwise directions have the strongest influence on the flow field
and the impact of these forces is largest on the streamwise
velocity. In some of the cases, the frequency response components
peak at non-zero temporal frequencies. This is distinct from
Newtonian fluids, where peaks are always observed at zero frequency,
suggesting that elasticity introduces additional timescales and
promotes development of flow patterns with smaller time constants
than in Newtonian fluids. We also find that the temporal
frequencies, corresponding to the peaks in the components of
frequency response, decrease with an increase in viscosity ratio and
show maxima with respect to the elasticity number. One of the most
important conclusions of this paper is the observation that
elasticity can lead to considerable energy amplification even when
inertial effects are weak; this energy amplification may then serve
as a route through which channel flows of Oldroyd-B fluids
transition to turbulence at low Reynolds
numbers~\citep{Larson2000,Groisman2000}.

Our presentation is organized as follows: in \S~\ref{sec.dff}, a
model for streamwise-constant channel flows of Oldroyd-B fluids with
external forcing is presented. In \S~\ref{sec.norms-inf}, a brief
summary of the notion of the spatio-temporal frequency response is
provided. In \S~\ref{sec.frRe}, an explicit scaling of the frequency
response components with the Reynolds number is given. In
\S~\ref{sec.FRkx0-PS} and \S~\ref{sec.Energy}, the effects of
elasticity number and viscosity ratio on power spectral and
steady-state energy densities are studied. The important findings
are summarized in \S~\ref{sec.conc}, and the detailed mathematical
derivations are relegated to the Appendix.

\section{The cross-sectional 2D/3C model}
    \label{sec.dff}

    \begin{figure}
    \centering
    \includegraphics[width=0.85\textwidth,clip]{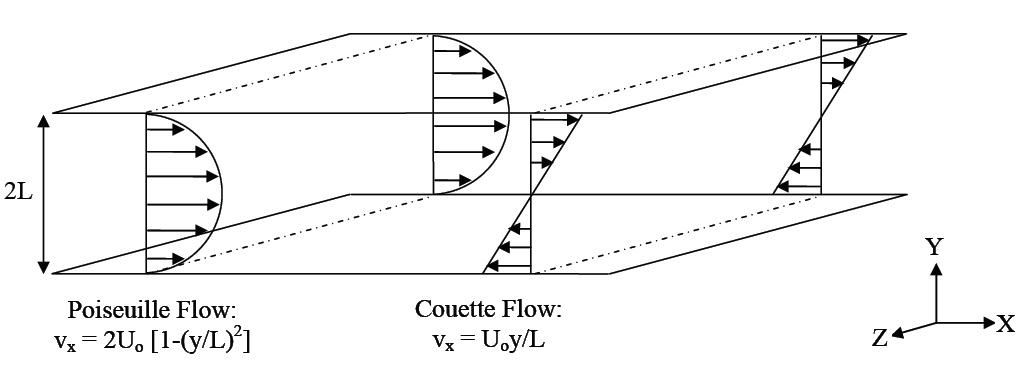}
    \caption{Schematic of the channel-flow geometry.}
    \label{Figure1}
    \end{figure}

A schematic of the channel-flow geometry is shown in
figure~\ref{Figure1}. The height of the channel is $2L$ and the
channel extends infinitely in the $x$- and $z$- directions. For
Couette flow, the top plate moves in the positive $x$-direction and
the bottom plate moves in the negative $x$-direction, each with
uniform velocity $U_o$. For Poiseuille flow, $2U_o$ is the
centerline velocity. We analyze the dynamical properties of the LNSE
for an Oldroyd-B fluid with spatially distributed and temporally
varying body-force fields. The parameters characterizing Oldroyd-B
fluids are:
    (a) the viscosity ratio, $\beta = \eta_s/(\eta_s + \eta_p)$,
where $\eta_s$ and $\eta_p$ are the solvent and polymer viscosities,
respectively;
    (b) the Reynolds number, $Re = U_o L/\nu = \rho U_o L/(\eta_s +
\eta_p)$, which represents the ratio of inertial to viscous forces,
where $\rho$ denotes fluid density;
    (c) the Weissenberg number, $\We = \lambda U_o/L$, which
characterizes the importance of the fluid relaxation time,
$\lambda$, with respect to the characteristic flow time, $L/U_o$.
    Another important parameter is the elasticity number,
$\mu=\We/Re=\lambda \nu/L^2$, which quantifies the ratio between the
fluid relaxation time, $\lambda$, and the vorticity diffusion time,
$L^2/\nu$.

Consider the dimensionless linearized momentum, continuity, and constitutive equations
    \beq
    \ba{rcl}
    \partial_t \mathbf{v}
    &\!\! = \!\!&
    -\; \mathbf{v} \cdot \mbox{\boldmath$\nabla$} \mathbf{\overline{v}}
    \;-\; \mathbf{\overline{v}} \cdot \mbox{\boldmath$\nabla$} \mathbf{v}
    \;+\; \dfrac{1}{Re} \Bigl( -\mbox{\boldmath$\nabla$}p  \;+\; \beta \nabla^2
    \mathbf{v}
    \;+\; (1 - \beta) \mbox{\boldmath $\nabla$} \cdot \mbox{\boldmath${\tau}$}
    \Bigr) \;+\; \mathbf{d},
    \\[-0.1cm]
    0
    &\!\! = \!\! &
    \mbox{\boldmath$\nabla$} \cdot \mathbf{v},
    \\[0.0cm]
    \partial_t \mbox{\boldmath${\tau}$}
    &\!\! = \!\!&
    \dfrac{1}{\We}\bigl( \mbox{\boldmath$\nabla$} \mathbf{v}
    \;+\; \left( \mbox{\boldmath$\nabla$} \mathbf{v} \right)^{T} \bigr)
    \;-\; \mathbf{v} \cdot
    \mbox{\boldmath$\nabla$}\mbox{\boldmath${\overline{\tau}}$}
    \;-\; \mathbf{\overline{v}} \cdot \mbox{\boldmath$\nabla$}
    \mbox{\boldmath$\tau$}
    \;+\; \mbox{\boldmath${\tau}$} \cdot \mbox{\boldmath$\nabla$}
    \mathbf{\overline{v}}
    \;+\; \mbox{\boldmath${\overline{\tau}}$} \cdot
    \mbox{\boldmath$\nabla$} \mathbf{v}
    \\
    &\!\! + \!\!& ( \mbox{\boldmath${\overline{\tau}}$}\cdot \mbox {\boldmath$\nabla$}
    \mathbf{v} )^{T}
    \;+\; ( \mbox{\boldmath${\tau}$}\cdot \mbox{\boldmath$\nabla$}
    \mathbf{\overline{v}})^{T}
    \;-\; \dfrac{\mbox{\boldmath${\tau}$}}{\We},
    \ea
    \label{eq.linear-original}
    \eeq
where $\mathbf{v} = \left[~u~~v~~w~\right]^{T}$ is the velocity
fluctuation vector, $p$ is the pressure fluctuation, and
$\mbox{\boldmath${\tau}$}$ is the polymer stress fluctuation. The
overbar denotes the base flow given by
    \beq
    \mathbf{\overline{v}}
    ~=~
    \left[~U(y)~~0~~0~\right]^T,
    ~~~
    \overline{\mbox{\boldmath{$\tau$}}}
    ~=~
    \left[
    \begin{array}{c c c}
    2 \We \left( U'(y) \right)^2 & U'(y) & 0
    \\[0.cm]
    U'(y) & 0 & 0
    \\[0.cm]
    0 & 0 & 0
    \end{array}
    \right],
    \non
    \eeq
with
    $
    U(y) \, = \, \{ y,~\mbox{Couette flow};
    $
    $
    1 - y^2,~\mbox{Poiseuille flow} \},
    $
and
    $U'(y) = \mrd U(y)/ \mrd y$.
A spatio-temporal body force is represented by $\bd =
\left[~d_1~~d_2~~d_3~\right]^{T}$, where $d_1, d_2$, and $d_3$ are
the body force fluctuations in the streamwise ($x$), wall-normal ($y$), and
spanwise ($z$) directions, respectively. These body forces can be
either deterministic or stochastic, and they serve as inputs into the
system of equations that governs evolution of
velocity and polymer stress fluctuations~\cite[][]{Mihailo2005}.
Our objective is to investigate their effect on the components of the
velocity field $\bv = \left[~u~~v~~w~\right]^{T}$.

In this paper, we confine our analysis to streamwise-constant
perturbations. In this special case, the model
for flow perturbations is usually referred to as the
two-dimensional, three-component (2D/3C) model~\cite[][]{reykas95}.
(2D indicates that the dynamics evolve in the cross-sectional ($y,z$)-plane,
and 3C indicates that velocity components in all three spatial
directions are considered.) The motivation for a thorough analysis
of this particular model is twofold: (a) a recent study by the
authors suggests that streamwise-constant (and nearly streamwise-constant)
perturbations in channel flows of Oldroyd-B fluids create
the largest contribution to the ensemble-average energy
density~\cite[][]{Nazish2008a}; and (b) for the 2D/3C model, an explicit
$Re$-dependence for the components of the frequency response can be
obtained, which clarifies the effectiveness (energy content) of
forcing (velocity) components.

The evolution model of the forced linearized system~(\ref{eq.linear-original}) with streamwise-constant perturbations ($k_x=0$) is obtained by a standard conversion~\citep{schhen01} to the wall-normal velocity/vorticity ($v,\omega_y$) formulation:
    \beq
    \ba{rcl}
    \pt \bpsi(y,k_z,t)
    & \!\! = \!\! &
    \left[ \, \bA (k_z) \bpsi(k_z,t) \, \right](y)
    \; + \;
    \left[ \, \bar{\bB}(k_z) \mathbf{d}(k_z,t) \, \right](y),
    \\[0.1cm]
    \bv (y,k_z,t)
    & \!\! = \!\! &
    \left[ \, \bC (k_z) \bpsi_1 (k_z,t) \, \right](y).
    \ea
    \label{eq.LNS}
    \eeq
Here, $k_z$ is the spanwise wavenumber,
    $
    \bpsi
    =
    \left[~\bpsi_1^T~~\bpsi_2^T~\right]^T,
    $
    $
    \bpsi_1
    =
    \left[~v~~\omega_y~\right]^T$,
and the components of polymer stress are
given by
    $
    \bpsi_2
    =
    \left[\,\tau_{xx}~\tau_{yy}~\tau_{zz}~\tau_{xy}~\tau_{xz}~\tau_{yz}\,\right]^{T}.
    $
The derivation of the evolution equation for $v$ requires elimination of the pressure from Eqs.~(\ref{eq.linear-original}). This is achieved by applying the divergence operator to the momentum equation and by combining the resulting equation with continuity. On the other hand, the equation for $\omega_y$ is derived by applying the curl operator to the momentum equation.

The operator $\bA$ in Eq.~(\ref{eq.LNS}) is referred to as the dynamical
generator of the linearized dynamics and it characterizes internal
properties of the LNSE (e.g., modal stability). The definition of
this operator for full three-dimensional fluctuations is provided
in~\citet{Nazish2008a}; the definition of components of this
operator suitable for frequency response analysis of the 2D/3C model
is given in Eqs.~(\ref{eq.lnse-2d3c})~and~(\ref{eq.F}) below.
We also note that operator $\bar{\bB}$ can be partitioned as
    $
    \bar{\bB}
    =
    \left[~\bB^T~~\mathbf{O}^T~\right]^{T},
    $
where $\bB$ describes how forcing enters into the Orr-Sommerfeld and
Squire equations of viscoelastic channel flows, and $\mathbf{O}$ is
a $6\times3$ matrix of null operators. On the other hand, operator
$\bC$ in Eq.~(\ref{eq.LNS}) contains information about a kinematic
relationship between $\bpsi_1$ and $\bv$.
These two operators are given by:
    \beq
    \bB
    ~=~
    \left[
    \begin{array}{ccc}
    0 & \bB_{2} & \bB_{3}
    \\[0.cm]
    \bB_{1} & 0 & 0 \end{array}
    \right],
    ~~~
    \bC
    ~=~
    \left[
    \begin{array}{cc}
    0 & \bC_{u} \\[0.cm]
    \bC_{v} & 0 \\[0.cm]
    \bC_{w} & 0 \end{array}
    \right],
    \non
    \eeq
    \beq
    \{
    \bB_1 = \mri k_z,
    \,
    \bB_2 = - k^2_z {\bDelta}^{-1},
    \,
    \bB_3 = - \mri k_z {\bDelta}^{-1} \py
    \},
    ~
    \{
    \bC_u = - (\mri/k_z),
    \,
    \bC_v = \bI,
    \,
    \bC_w = (\mri/k_z) \py
    \},
    \non
    \eeq
where $\mathrm{i}=\sqrt{-1}$, $\bI$ is the identity operator,
$\bDelta = \partial_{yy} - k_z^2 $ is a Laplacian with Dirichlet
boundary conditions, and $\bDelta^{-1}$ denotes the inverse of the
Laplacian. System~(\ref{eq.LNS}) is subject to the following
boundary conditions
    $
    \{
    v(\pm 1,k_z,t)
    =
    \py v(\pm 1,k_z,t)
    =
    \omega_y (\pm1,k_z,t)
    = 0
    \},
    $
which come from the no-slip and no-penetration requirements.
We note that no boundary conditions on
the polymer stresses are needed~\cite[][]{Nazish2008a}.

A coordinate transformation $\bphi = \bT \bpsi$ with
    $
    \{
    \phi_1 = v,
    $
    $
    \bphi_2 = \left[\,\tau_{yy}~\tau_{yz}~\tau_{zz}\,\right]^T,
    $
    $
    \phi_3 = \omega_y,
    $
    $
    \bphi_4
    =
    \left[\,\tau_{xy}~\tau_{xz}\,\right]^T,
    $
    $
    \phi_5
    =
    \tau_{xx}
    \}
    $
can be used to bring system~(\ref{eq.LNS}) into the following form:
    \begin{subequations}
    \label{eq.lnse-2d3c}
    \begin{align}
    \label{eq.lnse-2d3c-v}
    \pt \, \phi_1
    & ~=~
    \dfrac{\beta}{Re} \, \bF_{11} \, \phi_1
    ~+~
    \dfrac{1 \,-\, \beta}{Re} \, \bF_{12} \, \bphi_2
    ~+~
    \bB_2 \, d_2
    ~+~
    \bB_3 \, d_3,
    \\
    \label{eq.lnse-2d3c-tauv}
    \pt \, \bphi_2
    & ~=~
    - \dfrac{1}{\mu Re} \, \bphi_2
    ~+~
    \dfrac{1}{\mu Re} \, \bF_{21} \, \phi_1,
    \\
    \label{eq.lnse-2d3c-eta}
    \pt \, \phi_3
    & ~=~
    \dfrac{\beta}{Re} \, \bF_{33} \, \phi_3
    ~+~
    \bF_{31} \, \phi_1
    ~+~
    \dfrac{1 \,-\, \beta}{Re} \, \bF_{34} \, \bphi_4
    ~+~
    \bB_1 \, d_1,
    \\
    \label{eq.lnse-2d3c-taueta}
    \pt \, \bphi_4
    & ~=~
    - \dfrac{1}{\mu Re} \, \bphi_4
    ~+~
    \bF_{41} \, \phi_1
    ~+~
    \bF_{42} \, \bphi_2
    ~+~
    \dfrac{1}{\mu Re} \, \bF_{43} \, \phi_3,
    \\
    \label{eq.lnse-2d3c-tau11}
    \pt \, \phi_5
    & ~=~
    - \dfrac{1}{\mu Re} \, \phi_5
    ~-~
    {\mu Re} \, \bF_{51} \, \phi_1
    ~+~
    \bF_{53} \, \phi_3
    ~+~
    \bF_{54} \, \bphi_4,
    \\[0.1cm]
    \label{eq.lnse-2d3c-out}
    \left[
    \ba{c}
    {u} \\[0.cm]
    {v} \\[0.cm]
    {w}
    \ea
    \right]
    & ~=~
    \left[
    \begin{array}{cc}
    0 & \bC_{u} \\[0.cm]
    \bC_{v} & 0 \\[0.cm]
    \bC_{w} & 0 \end{array}
    \right]
    \left[
    \ba{c}
    {\phi_1} \\[0.cm]
    {\phi_3}
    \ea
    \right],
    \end{align}
    \end{subequations}
where the $\bF$-operators are given by:
    \beq
    \ba{c}
    \bF_{11}
    \; = \;
    \bDelta^{-1} \bDelta^{2},
    ~~
    \bF_{33}
    \; = \;
    \bDelta,
    ~~
    \bF_{31}
    \; =  \;
    - \,\mri k_z U'(y),
    \\[0.15cm]
    \bF_{12}
    \; =  \;
    \bDelta^{-1}
    \left[
    \ba{ccc}
    -k_z^2 \py
    &
    - \mri k_z
    \left(
    \pyy \, + \, k_z^2
    \right)
    &
    k_z^2 \py
    \ea
    \right],
    ~~
    \bF_{34}
    \; = \;
    \left[
    \ba{cc}
    \mri k_z \py
    &
    - k_z^2
    \ea
    \right],
    \\[0.15cm]
    \bF_{21}
    \; = \;
    \left[
    \ba{ccc}
    2 \py
    &
    (\mri/k_z)
    \left(
    \pyy \, + \, k_z^2
    \right)
    &
    -2 \py
    \ea
    \right]^T,
    ~~
    \bF_{43}
    \; = \;
    - \, (1/k_z^2)
    \bF_{34}^T,
    \\[0.15cm]
    \bF_{41}
    \; = \;
    \left[
    \ba{c}
    U'(y) \py  - U''(y)
    \\[0.cm]
    (\mri/k_z) U'(y) \pyy
    \ea
    \right],
    ~~
    \bF_{42}
    \; = \;
    \left[
    \ba{ccc}
    U'(y) & 0 & 0
    \\[0.cm]
    0 & U'(y) & 0
    \ea
    \right],
    \\[0.25cm]
    \bF_{51}
    \; = \;
    4 U'(y) U''(y),
    ~~
    \bF_{53}
    \; = \;
    - \, (2 \mri/k_z) U'(y) \py,
    ~~
    \bF_{54}
    \; = \;
    \left[
    \ba{cc}
    2 U'(y)
    &
    0
    \ea
    \right].
    \ea
    \label{eq.F}
    \eeq
Here, $\bDelta^2 = \partial_{yyyy} - 2 k_z^2\partial_{yy} + k_z^4$ with
both Dirichlet and Neumann boundary conditions.

The system of equations~(\ref{eq.lnse-2d3c}) is in a form suitable
for the analysis performed in \S~\ref{sec.frRe} where an explicit
characterization of the Reynolds-number dependence for the
components of the frequency response of system~(\ref{eq.LNS}) is provided.
It is noteworthy that for the 2D/3C model there is no coupling from
    $
    \phi_5
    \, = \,
    \tau_{xx}
    $
to the equations for the other flow-field components
in~(\ref{eq.lnse-2d3c}); in particular, this demonstrates that
evolution of $\tau_{xx}$ at $k_x = 0$ does not influence evolution
of $u$, $v$, and $w$. We also note a one-way coupling from
Eqs.~(\ref{eq.lnse-2d3c-v})~and~(\ref{eq.lnse-2d3c-tauv}) to
Eqs.~(\ref{eq.lnse-2d3c-eta})~and~(\ref{eq.lnse-2d3c-taueta}); this
indicates that the dynamical properties of
    $
    \phi_3 = \omega_y
    $
and
    $
    \bphi_4
    =
    \left[\,\tau_{xy}~\tau_{xz}\,\right]^T
    $
are influenced by
    $
    \phi_1 = v
    $
and
    $
    \bphi_2 = \left[\,\tau_{yy}~\tau_{yz}~\tau_{zz}\,\right]^T
    $
but not vice-versa.

\section{Frequency responses for streamwise-constant perturbations}
    \label{sec.norms-inf}

Frequency response represents a cornerstone of input-output
analysis of linear dynamical systems~\cite[][]{Zhou1996}. The utility
of input-output analysis in understanding early stages of transition
in wall-bounded shear flows of Newtonian fluids is by now well
documented; we refer the reader to a recent review article
by~\citet{Schmid2007} for more information. It turns out that
the input-output approach also reveals important facets of transitional
dynamics in channel flows of Oldroyd-B
fluids~\cite[][]{Nazish2008a}.

To provide a self-contained treatment, we next present a brief
summary of the notion of the spatio-temporal frequency response of
the streamwise-constant LNSE with forcing; we invite the reader to
see~\citet{Mihailo2005} for additional details. The spatio-temporal
frequency response of system~(\ref{eq.LNS}) is given by
    \beq
    \bH (k_z,\omega)
    ~=~
    \bC (k_z)
    ( \mri \omega \bI -  \bA (k_z))^{-1}  \bar{\bB} (k_z),
    \non
    \eeq
where $\omega$ denotes the temporal frequency. The frequency
response is obtained directly from the Fourier symbols of the
operators in Eq.~(\ref{eq.LNS}), and for any pair ($k_z,\omega$) it
represents an operator (in $y$) that maps the forcing field into the
velocity field.

The frequency response of a system with a stable generator $\bA$
describes the steady-state response to harmonic input signals across
temporal and spatial frequencies. Since $\bH$ is an operator valued function of two independent variables $k_z$ and $\omega$, there are a variety of ways to
visualize its properties. In this paper, we study the Hilbert--Schmidt norm of $\bH$
    \beq
    \Pi (k_z,\omega)
    \; = \;
    \trace
    \left(
    \bH (k_z,\omega) \bH^{*} (k_z,\omega)
    \right),
    \label{eq.HS}
    \eeq
where $ \bH^*$ represents the adjoint of operator $\bH$. For any pair $(k_z,\omega)$, the Hilbert--Schmidt norm
quantifies the power spectral density of the velocity field in the
LNSE subject to harmonic (in $z$) white, unit variance, temporally
stationary, stochastic (in $y$ and $t$) body forcing.
Furthermore, the temporal-average of the power spectral density of
$\bH$ yields the \mbox{so-called} $H_2$ norm
of system~(\ref{eq.LNS})~\citep{Zhou1996}
    \beq
    E (k_z)
    \; = \;
    \dfrac{1}{2 \pi}
    \int_{-\infty}^{\infty}
    \Pi (k_z,\omega)
    \, \mrd \omega.
    \non
    \eeq
The frequency responses of viscoelastic channel flows (as a function
of $k_x$ and $k_z$) are quantified in~\cite{Nazish2008a} in terms of
the $H_2$ norm. We note that at any $k_z$, the $H_2$ norm determines the energy (variance) amplification of harmonic (in $z$) stochastic (in $y$ and $t$)
disturbances~\cite[][]{Farrell1993,Bamieh2001,Mihailo2005}. This
quantity is also known as the {\em ensemble-average energy
density\/} of the statistical steady-state~\cite[][]{Farrell1993}, and it is hereafter referred to as the (steady-state) energy density (or energy amplification).

We finally note that the frequency response of system~(\ref{eq.LNS}), $\bv
= \bH \bd$, has the following $3 \times 3$ block-decomposition:
    \beq
    \left[
    \ba{c}
    u \\[0cm]
    v \\[0.cm]
    w
    \ea
    \right]
    =
    \left[
    \ba{ccc}
    \bH_{u1}(k_z,\omega;Re,\beta,\mu) & \bH_{u2}(k_z,\omega;Re,\beta,\mu) & \bH_{u3}(k_z,\omega;Re,\beta,\mu)\\[0.cm]
    \bH_{v1}(k_z,\omega;Re,\beta,\mu) & \bH_{v2}(k_z,\omega;Re,\beta,\mu) & \bH_{v3}(k_z,\omega;Re,\beta,\mu)\\[0.cm]
    \bH_{w1}(k_z,\omega;Re,\beta,\mu) & \bH_{w2}(k_z,\omega;Re,\beta,\mu) & \bH_{w3}(k_z,\omega;Re,\beta,\mu)
    \ea
    \right ]
    \left[
    \ba{c}
    d_1 \\[0.cm]
    d_2 \\[0.cm]
    d_3
    \ea
    \right],
    \label{eq.FR}
    \eeq
which is suitable for uncovering the effectiveness (energy content)
of forcing (velocity) components. In this representation, $\bH_{rj}$
denotes the frequency response operator from $d_j$ to $r$, with $\{j
= 1,2,3;$ $r = u,v,w \}$. Our notation suggests that $\!-\!$ in
addition to the spanwise wavenumber $k_z$ and the temporal frequency
$\omega$ $\!-\!$ each component of $\bH$ also depends on the
Reynolds number $Re$, the viscosity ratio $\beta$, and the
elasticity number $\mu$.

\section{Dependence of frequency responses on the Reynolds number}
    \label{sec.frRe}

In this section, we study how the power spectral densities and the
steady-state energy densities scale with $Re$ for each of the
components of the frequency response~(\ref{eq.FR}). Furthermore, the
square-additive property of these two quantities is used to
determine the aggregate effect of forces in all three spatial
directions $\bd = \left[~d_1~~d_2~~d_3~\right]^{T}$ on all three
velocity components $\bv = \left[~u~~v~~w~\right]^{T}$. We
analytically establish that the frequency responses from both wall-normal and spanwise forces to streamwise velocity scale as $Re^2$, while the frequency responses of all other components in Eq.~(\ref{eq.FR}) scale at most~as~$Re$. This
extends the Newtonian-fluid results~\citep{mj-phd04,Mihailo2005} to
channel flows of Oldroyd-B fluids.

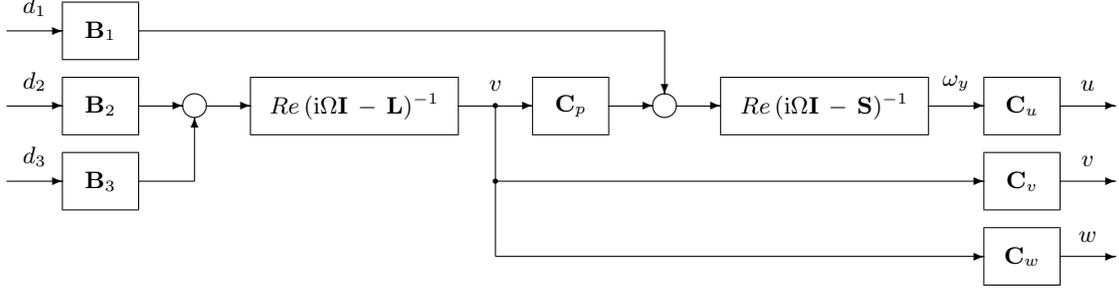
\begin{figure}
    \centering
    \begin{picture}(14.75,3.75)(0,0)
   % Block diagram for the linearized NS equations

    %first part - dv
    \put(0,2.375){\vector(1,0){0.75}}
    \put(0.375,2.575){\makebox(0,0)[b]{$ d_2 $}}
    \put(0.75,2){\framebox(1,0.75){$\bB_2$}}
    \put(1.75,2.375){\vector(1,0){0.6}}
    \put(2.5,2.375){\circle{0.3}}

    %second part - orr-sommerfeld
    \put(2.65,2.375){\vector(1,0){0.6}}
    \put(3.25,2){\framebox(2.75,0.75){$ Re \, (\mathrm{i}\Omega \mathbf{I} \, - \, \bL)^{-1} $}}
    \put(6,2.375){\line(1,0){0.5}}
    \put(6.5,2.375){\circle*{0.08}}
    \put(6.5,2.375){\vector(1,0){0.5}}
    %third part - coupling
    \put(7,2){\framebox(1,0.75){$\bC_p$}}
    \put(8,2.375){\vector(1,0){0.6}}
    \put(8.75,2.375){\circle{0.3}}

    %fourth part - Squire
    \put(8.9,2.375){\vector(1,0){0.6}}
    \put(9.5,2){\framebox(2.75,0.75){$ Re \, (\mathrm{i} \Omega \mathbf{I} \, - \, \bS)^{-1} $}}
    \put(12.25,2.375){\vector(1,0){0.75}}
    \put(12.625,2.575){\makebox(0,0)[b]{$ \omega_y $}}

    % Map to u
    \put(13,2){\framebox(1,0.75){$\bC_u $}}
    \put(14,2.375){\vector(1,0){0.75}}
    \put(14.375,2.575){\makebox(0,0)[b]{$ u $}}
  
    %part 5 - du to v
    \put(0,3.375){\vector(1,0){0.75}}
    \put(0.375,3.575){\makebox(0,0)[b]{$ d_1 $}}
    \put(0.75,3){\framebox(1,0.75){$\bB_1$}}
    \put(1.75,3.375){\line(1,0){7}}
    \put(8.75,3.375){\vector(0,-1){0.85}}

    %part 6 - dw to v
    \put(0,1.375){\vector(1,0){0.75}}
    \put(0.375,1.575){\makebox(0,0)[b]{$ d_3 $}}
    \put(0.75,1){\framebox(1,0.75){$\bB_3$}}
    \put(1.75,1.375){\line(1,0){0.75}}
    \put(2.5,1.375){\vector(0,1){0.85}}

    %part 7 - du to w
    \put(6.5,2.375){\line(0,-1){1}}
    \put(6.5,1.375){\circle*{0.08}}
    \put(6.5,1.375){\vector(1,0){6.5}}
    \put(13,1){\framebox(1,0.75){$\bC_v$}}
    \put(14,1.375){\vector(1,0){0.75}}
    \put(14.375,1.575){\makebox(0,0)[b]{$v$}}

    %part 8 - dw to w
    \put(6.5,1.375){\line(0,-1){1}}
    \put(6.5,0.375){\vector(1,0){6.5}}
    \put(13,0){\framebox(1,0.75){$ \bC_w$}}
    \put(14,0.375){\vector(1,0){0.75}}
    \put(14.375,0.575){\makebox(0,0)[b]{$w$}}
    
    %part 9 - v
    \put(6.5,2.575){\makebox(0,0)[b]{$v$}}
\end{picture}
    \caption{Block diagram representation of the streamwise constant LNS
    system~(\ref{eq.LNS}).}
    \label{fig.Block}
    \end{figure}

Application of the temporal Fourier transform to Eq.~(\ref{eq.lnse-2d3c}) facilitates elimination of polymer stresses from the 2D/3C model (see Appendix~\ref{append.FR} for details). This leads to an equivalent representation of system~(\ref{eq.lnse-2d3c}) in terms of its block diagram, which is shown in figure~\ref{fig.Block} with $\Omega = \omega Re$. From this
block diagram, it follows that operator $\bH(k_z,\omega;Re,\beta,\mu)$ in Eq.~(\ref{eq.FR}) can be expressed as
    \beq
    \left[
    \ba{c}
    u \\[0.cm]
    v \\[0.cm]
    w
    \ea
    \right]
    =
    \left[
    \ba{cll}
    Re \, \bar{\bH}_{u1}(k_z,\Omega;\beta,\mu) & Re^2 \, \bar{\bH}_{u2}(k_z,\Omega;\beta,\mu) & Re^2 \, \bar{\bH}_{u3}(k_z,\Omega;\beta,\mu)\\[0.cm]
    0 & Re \, \bar{\bH}_{v2}(k_z,\Omega;\beta,\mu) & Re \, \bar{\bH}_{v3}(k_z,\Omega;\beta,\mu)\\[0.cm]
    0 & Re \, \bar{\bH}_{w2}(k_z,\Omega;\beta,\mu) & Re \, \bar{\bH}_{w3}(k_z,\Omega;\beta,\mu)
    \ea
    \right ]
    \left[
    \ba{c}
    d_1 \\[0.cm]
    d_2 \\[0.cm]
    d_3
    \ea
    \right],
    \label{eq.lnse-io}
    \eeq
where the Reynolds-number-independent operators $\bar{\bH}$ are
given by:
    \beq
    \ba{rcl}
    \bar{\bH}_{u1}(k_z,\Omega;\beta,\mu)
    & \! = \! &
    \bC_u (\mri \Omega \bI \,-\, \bS)^{-1} \bB_1,
    \\[0.15cm]
    \bar{\bH}_{u j}(k_z,\Omega;\beta,\mu)
    & \! = \! &
    \bC_u (\mri \Omega  \bI \,-\, \bS)^{-1} \bC_p (\mri \Omega \bI \,-\, \bL)^{-1} \bB_j,
    ~~~
    j = 2,3,
    \\[0.15cm]
    \bar{\bH}_{r j}(k_z,\Omega;\beta,\mu)
    & \! = \! &
    \bC_r (\mri \Omega \bI \,-\, \bL)^{-1} \bB_j,
    ~~~
    \{r = v,w; \; j = 2,3 \}.
    \ea
    \non
    \eeq
Operators $\bL$, $\bS$, and $\bC_p$ are defined by:
    \beq
    \ba{c}
    \bL
    \; = \;
    \dfrac{1 \, + \, \mri \beta \mu \Omega}{1 \, + \, \mri \mu\Omega}
    \bDelta^{-1} \bDelta^{2},
    ~~~
    \bS
    \; = \;
    \dfrac{1 \,+\, \mri \beta \mu \Omega}{1 \,+\, \mri \mu \Omega} \bDelta,
    ~~~
    \bC_p
    \; = \;
    \bC_{p1}
    \, + \,
    \dfrac{\mu}{(1 \,+\, \mri \mu \Omega)^2}
    \bC_{p2},
    \\[0.25cm]
    \bC_{p1}
    \; = \;
    - \mri k_z
    U'(y),
    ~~~
    \bC_{p2}
    \; = \;
    \mri k_z
    (1 \,-\, \beta)
    \left(
    U'(y) \bDelta \, + \, 2 U''(y) \py
    \right).
    \ea
    \non
    \eeq
In the limit $\beta \rightarrow 1$, these operators simplify to the
familiar Orr-Sommerfeld ($\bDelta^{-1} \bDelta^{2}$), Squire
($\bDelta$), and coupling ($- \mri k_z U'(y)$) operators in the
streamwise-constant LNSE of Newtonian fluids with $Re = 1$. We note
that even though viscoelastic effects modify some of the operators
in figure~\ref{fig.Block}, there is a striking similarity between
block-diagram representations of the 2D/3C models of non-Newtonian
and Newtonian fluids~\cite[][]{Mihailo2005}. In particular,
figure~\ref{fig.Block} shows that the frequency responses from $d_2$
and $d_3$ to $u$ scale as $Re^2$, whereas the responses from all
other forcing components to other velocity components scale linearly
with $Re$. It should be noted that the frequency responses of
Newtonian fluids at $k_x \, = \, 0$ show same scaling with
$Re$~\cite[][]{mj-phd04,Mihailo2005}. The coupling operator, $\bC_p$, is
crucial for the $Re^2$-scaling. In Newtonian fluids $\bC_p$
corresponds to the vortex tilting term, $\bC_{p1}$; in viscoelastic
fluids $\bC_p$ also contains an additional term, $\bC_{p2}$, that
captures the coupling from the wall-normal velocity to the
wall-normal vorticity due to the work done by the polymer stresses
on the flow. In the absence of the coupling operator, all the
components of $\bH (\omega,k_z;Re,\beta,\mu)$ scale at most linearly
with $Re$. Figure~\ref{fig.Block} also suggests that for the 2D/3C
model, streamwise forcing does not influence the wall-normal and
spanwise velocities, which is in agreement with Newtonian-fluid
results~\citep{Mihailo2005}. As noted in Appendix~\ref{append.FR},
$\bC_{p2}$ has its origin in the term involving polymer stress
fluctuations and gradients in the base velocity profile. This
would produce polymer stretching, and could be interpreted as
giving rise to an effective lift-up mechanism~\citep{Landahl1975}.

The $Re$-scaling for the power spectral densities of operators $\bH_{r
j}(k_z,\omega;Re,\beta,\mu)$ in streamwise-constant Poiseuille and Couette flows of Oldroyd-B fluids follows directly from Eqs.~(\ref{eq.HS}) and~(\ref{eq.lnse-io}) and linearity of the trace operator:
    \beq
    \begin{aligned}
    &
    \left[
    \begin{array}{ccc}
    \Pi_{u 1} (k_z,\omega;Re,\beta,\mu)
    &
    \Pi_{u 2} (k_z,\omega;Re,\beta,\mu)
    &
    \Pi_{u 3} (k_z,\omega;Re,\beta,\mu)
    \\[0.cm]
    \Pi_{v 1} (k_z,\omega;Re,\beta,\mu)
    &
    \Pi_{v 2} (k_z,\omega;Re,\beta,\mu)
    &
    \Pi_{v 3} (k_z,\omega;Re,\beta,\mu)
    \\[0.cm]
    \Pi_{w 1} (k_z,\omega;Re,\beta,\mu)
    &
    \Pi_{w 2} (k_z,\omega;Re,\beta,\mu)
    &
    \Pi_{w 3} (k_z,\omega;Re,\beta,\mu)
    \end{array}
    \right]
    \\
    =~
    &
    \left[ \begin{array} {ccc}
    {\bPi_{u 1}(k_z,\Omega;\beta,\mu) Re^2}
    &
    {\bPi_{u 2}(k_z,\Omega;\beta,\mu) Re^4}
    &
    {\bPi_{u 3}(k_z,\Omega;\beta,\mu)Re^4}
    \\[0.cm]
    {0}
    &
    {\bPi_{v 2}(k_z,\Omega;\beta,\mu) Re^2}
    &
    {\bPi_{v 3}(k_z,\Omega;\beta,\mu) Re^2}
    \\[0.cm]
    {0}
    &
    {\bPi_{w 2}(k_z,\Omega;\beta,\mu) Re^2}
    &
    {\bPi_{w 3}(k_z,\Omega;\beta,\mu) Re^2}
    \end{array}
    \right],
    \end{aligned}
    \label{eq.hs-component}
    \end{equation}
where $\bPi_{rj}$ are the power spectral densities of the Reynolds-number-independent operators $\bar{\bH}_{rj} (k_z,\Omega;\beta,\mu)$, with  $\Omega \, = \, \omega Re$. Furthermore, the power spectral density
of operator $\bH(k_z,\omega;Re,\beta,\mu)$, $\bv \, = \, \bH \bd$,
is given by
    \beq
    \Pi (k_z,\omega;Re,\beta,\mu)
    \; = \;
    \bPi_a (k_z,\Omega;\beta,\mu) Re^2
    \; + \;
    \bPi_b (k_z,\Omega;\beta,\mu) Re^4,
    \non
    \eeq
where
    $
    \bPi_a
    \, = \,
    \bPi_{u 1} \, + \, \bPi_{v 2} \, + \, \bPi_{v 3} \, + \, \bPi_{w 2} \, + \, \bPi_{w 3}
    $
and
    $
    \bPi_b
    \, = \,
    \bPi_{u 2} \, + \, \bPi_{u 3}.
    $

Several important observations can be made about
Eq.~(\ref{eq.hs-component}) without doing any detailed
calculations.  First,
the power spectral densities of operators $\bH_{u2}$ and $\bH_{u3}$ scale as $Re^4$; in all other cases they scale at most as $Re^2$.   This
illustrates the dominance of the streamwise velocity perturbations
and the forces in the wall-normal and spanwise directions in
high-Reynolds-number channel flows of streamwise constant Oldroyd-B fluids.
Second,
apart from $\bPi_{u 2}(k_z,\Omega;\beta,\mu)$ and $\bPi_{u 3}(k_z,\Omega;\beta,\mu)$, the other power spectral densities
in Eq.~(\ref{eq.hs-component}) do not depend on the base velocity and stresses.  These two power spectral densities depend on the coupling operator, $\bC_p$, and thus their values differ in Poiseuille and Couette flows.  Third,
power spectral densities $\bPi_{r j}(k_z,\Omega;\beta,\mu)$ do not depend on the Reynolds number.  Thus, $Re$ only affects the magnitudes of $\Pi_{r j}(k_z,\omega;Re,\beta,\mu)$, and regions
of temporal frequencies $\omega$ where these power spectral densities peak.
As $Re$ increases, these $\omega$-regions shrink as $1/Re$.
Therefore, for high-Reynolds-number channel flows of Oldroyd-B
fluids, the influence of small temporal frequencies dominates the
evolution of the velocity perturbations, suggesting preeminence of
the effects in fluids with relatively large time constants. It is
noteworthy that elasticity shifts temporal frequencies where
$\bPi_{r j}$ peak to higher values, which is discussed in
detail in \S~\ref{sec.FRkx0-PS}.  For additional details
concerning these points, we refer the reader to~\citet{Nazish}.

We next exploit the above results to establish the Reynolds-number dependence of steady-state energy densities for different components of frequency response operator~(\ref{eq.FR}). For example, $E_{u 2} (k_z;Re,\beta,\mu)$ is determined by
    \beq
    \ba{rll}
    E_{u 2} (k_z;Re,\beta,\mu)
    & \! = \! &
    \ds{\dfrac{1}{2 \pi} \int_{-\infty}^{\infty}}
    \Pi_{u 2} (\omega,k_z;Re,\beta,\mu)
    \, \mrd \omega
    \\[0.35cm]
    & \! = \! &
    \ds{\dfrac{Re^4}{2 \pi} \int_{-\infty}^{\infty}}
    \bPi_{u 2} (\Omega,k_z;\beta,\mu)
    \, \mrd \omega
    \\[0.35cm]
    & \! = \! &
    \ds{\dfrac{Re^3}{2 \pi} \int_{-\infty}^{\infty}}
    \bPi_{u 2} (\Omega,k_z;\beta,\mu)  \, \mrd \Omega
    \\[0.35cm]
    & \! =: \! &
    Re^3 g_{u 2} (k_z;\beta,\mu).
    \ea
    \non
    \eeq
A similar procedure can be used to determine the steady-state energy
densities of all other components of operator $\bH$ in Eq.~(\ref{eq.FR}), which yields:
    \beq
    \begin{aligned}
    &
    \left[
    \begin{array}{lll}
    E_{u1}  (k_z;Re,\beta,\mu)
    &
    E_{u2}  (k_z;Re,\beta,\mu)
    &
    E_{u3}  (k_z;Re,\beta,\mu)
    \\[0.cm]
    E_{v1}  (k_z;Re,\beta,\mu)
    &
    E_{v2}  (k_z;Re,\beta,\mu)
    &
    E_{v3}  (k_z;Re,\beta,\mu)
    \\[0.cm]
    E_{w1}  (k_z;Re,\beta,\mu)
    &
    E_{w2}  (k_z;Re,\beta,\mu)
    &
    E_{w3}  (k_z;Re,\beta,\mu)
    \end{array}   \right]
    \\
    =~
    &
    \left[ \begin{array}{ccc}
    {f_{u1}(k_z;\beta,\mu) Re}
    &
    {g_{u2}(k_z;\beta,\mu) Re^3}
    &
    {g_{u3}(k_z;\beta,\mu) Re^3}
    \\[0.cm]
    {0}
    &
    {f_{v2}(k_z;\beta,\mu) Re}
    &
    {f_{v3}(k_z;\beta,\mu) Re}
    \\[0.cm]
    {0}
    &
    {f_{w2}(k_z;\beta,\mu) Re}
    &
    {f_{w3}(k_z;\beta,\mu) Re}
    \end{array}
    \right],
    \end{aligned}
    \non
    \eeq
where $f_{rj}$ and $g_{rj}$ are functions independent of $Re$.
Furthermore, the steady-state energy density of operator
$\bH(k_z,\omega;Re,\beta,\mu)$, $\bv \, = \, \bH \bd$, is given by
    \beq
    E (k_z;Re,\beta,\mu)
    \; = \;
    f (k_z;\beta,\mu) Re
    \; + \;
    g (k_z;\beta,\mu) Re^3,
    \label{eq.E-total}
    \eeq
where
    $
    f
    \, = \,
    f_{u 1} \, + \, f_{v 2} \, + \, f_{v 3} \, + \, f_{w 2} \, + \, f_{w 3}
    $
and
    $
    g
    \, = \,
    g_{u 2} \, + \, g_{u 3}.
    $

We conclude that energy amplification from both spanwise and
wall-normal forcing to streamwise velocity is $O(Re^3)$, while
energy amplification for all other components is $O(Re)$.

\section{Parametric study of power spectral densities}
    \label{sec.FRkx0-PS}

In \S~\ref{sec.frRe}, we derived an explicit dependence for each
component of the frequency response operator~(\ref{eq.FR}) on the
Reynolds number. Here, we investigate the effect of $\beta$ and
$\mu$ on the ($\Omega$, $k_z$)-parameterized plots of power spectral
densities $\bPi_{rj}$, $\{ r= u, v, w;$ $j = 1,2,3 \}$, by setting
$Re = 1$ in Eq.~(\ref{eq.hs-component}). In all the plots presented in
this section, $100 \times 90$ logarithmically spaced grid points are
used in the $(\Omega,k_z)$-plane. The temporal frequency and
spanwise wavenumber are varied between $0.01$ and $25.11$ ($\Omega$)
and $0.1$ and $15.84$ ($k_z$), respectively. The Reynolds-number-independent power spectral densities $\bPi_{rj} (k_z,\Omega;\beta,\mu)$ in Eq.~(\ref{eq.hs-component}) can either be numerically determined from
a finite-dimensional approximation of the underlying operators or they
can be computed using the method developed by~\cite{Mihailo2006}.
For numerical approximation, we use a Chebyshev collocation
technique~\citep{Reddy2000}; between $30$ and $50$ collocation
points were found to be sufficient to obtain accurate results. In
Couette flow, the power spectral densities can be computed more
efficiently using the method developed by~\cite{Mihailo2006}. This
can be accomplished by expressing each component of the frequency
response operator in a different form, known as the two-point
boundary value state-space realization~\citep{Nazish}. In Couette
flow, we used both methods for evaluating $\bPi_{rj}$; the
results agreed with each other, suggesting accuracy of our
computations.

\begin{figure}
    \begin{center}
    \begin{tabular}{cc}
    $\bPi_{u 1}(k_z,\Omega;\beta,\mu)$ & $\bPi_{v 2}(k_z,\Omega;\beta,\mu)$
    \\
    \includegraphics[width=0.5\textwidth]{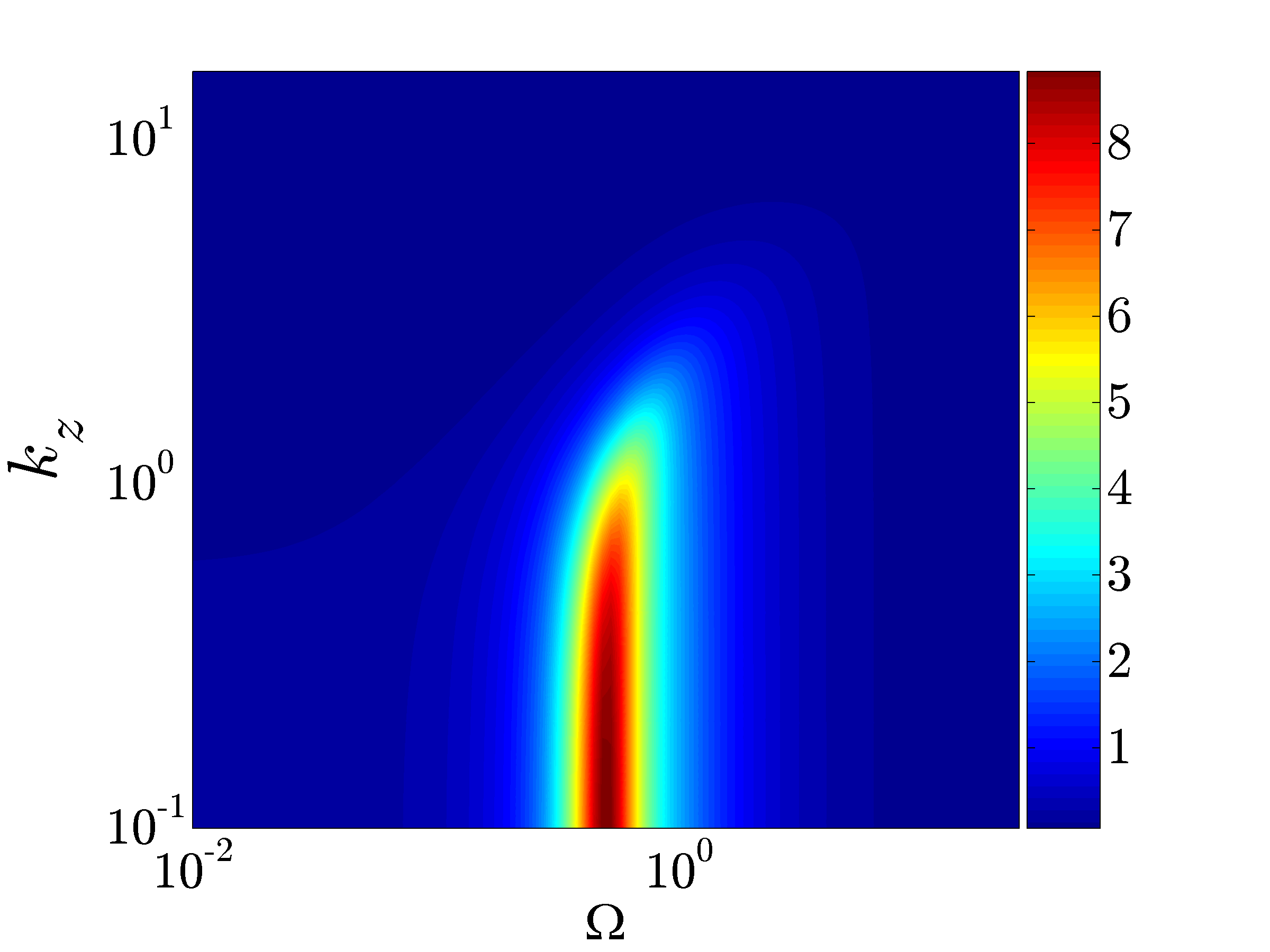}
    &
    \includegraphics[width=0.5\textwidth]{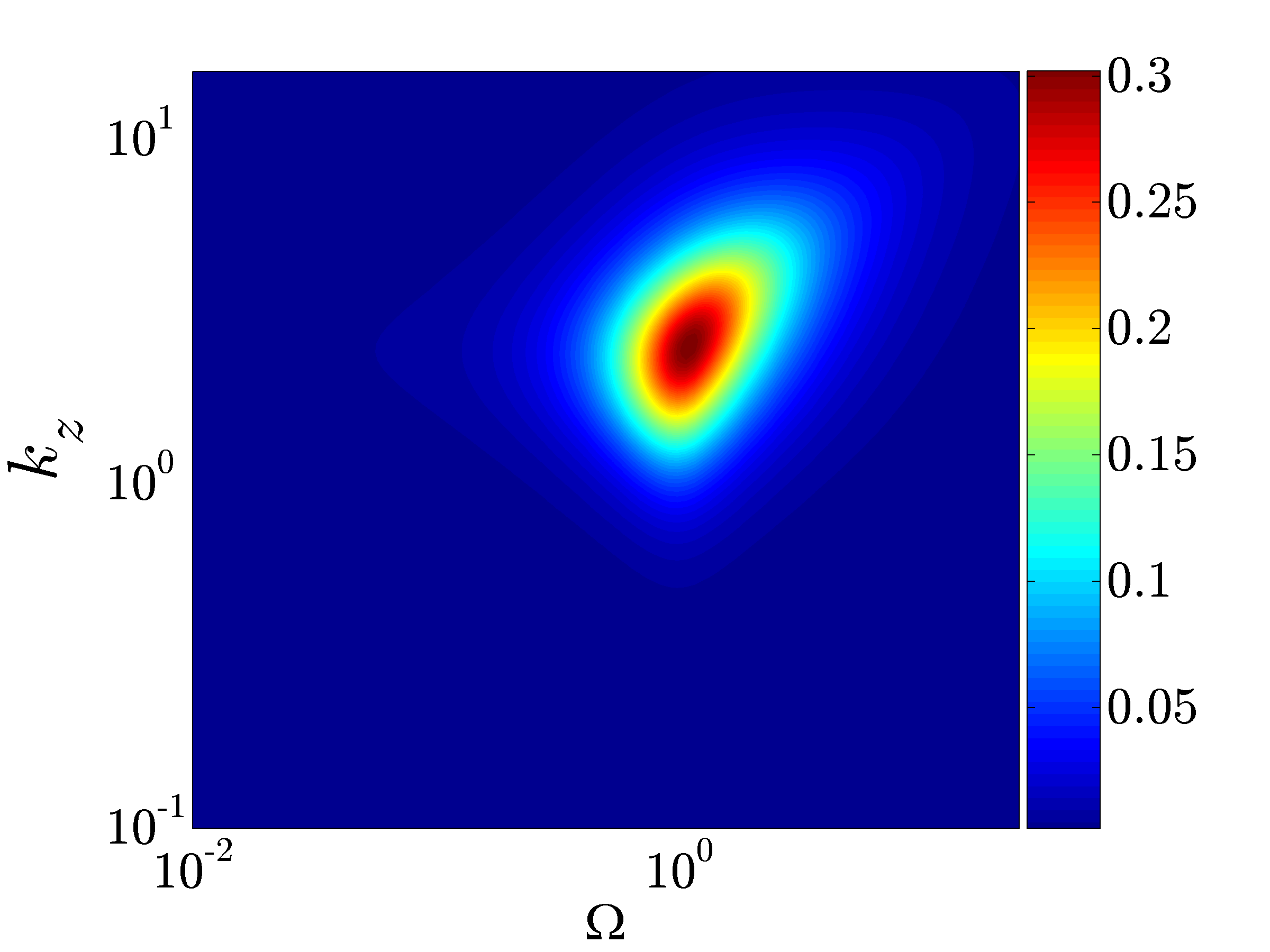}
    \\
    $\bPi_{v 3} (k_z,\Omega;\beta,\mu)$
    &
    $\bPi_{w 3} (k_z,\Omega;\beta,\mu)$
    \\
    \includegraphics[width=0.5\textwidth]{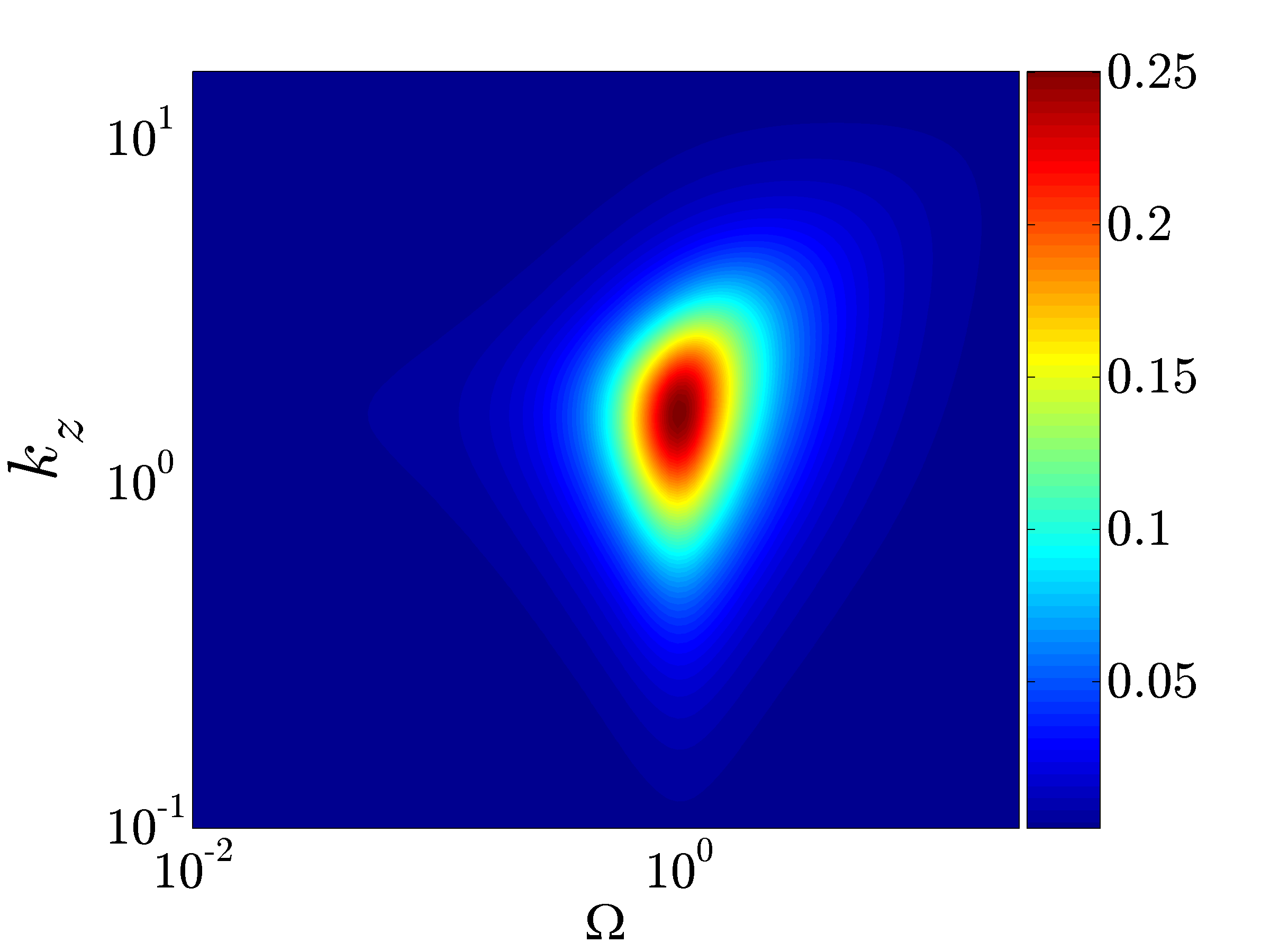}
    &
    \includegraphics[width=0.5\textwidth]{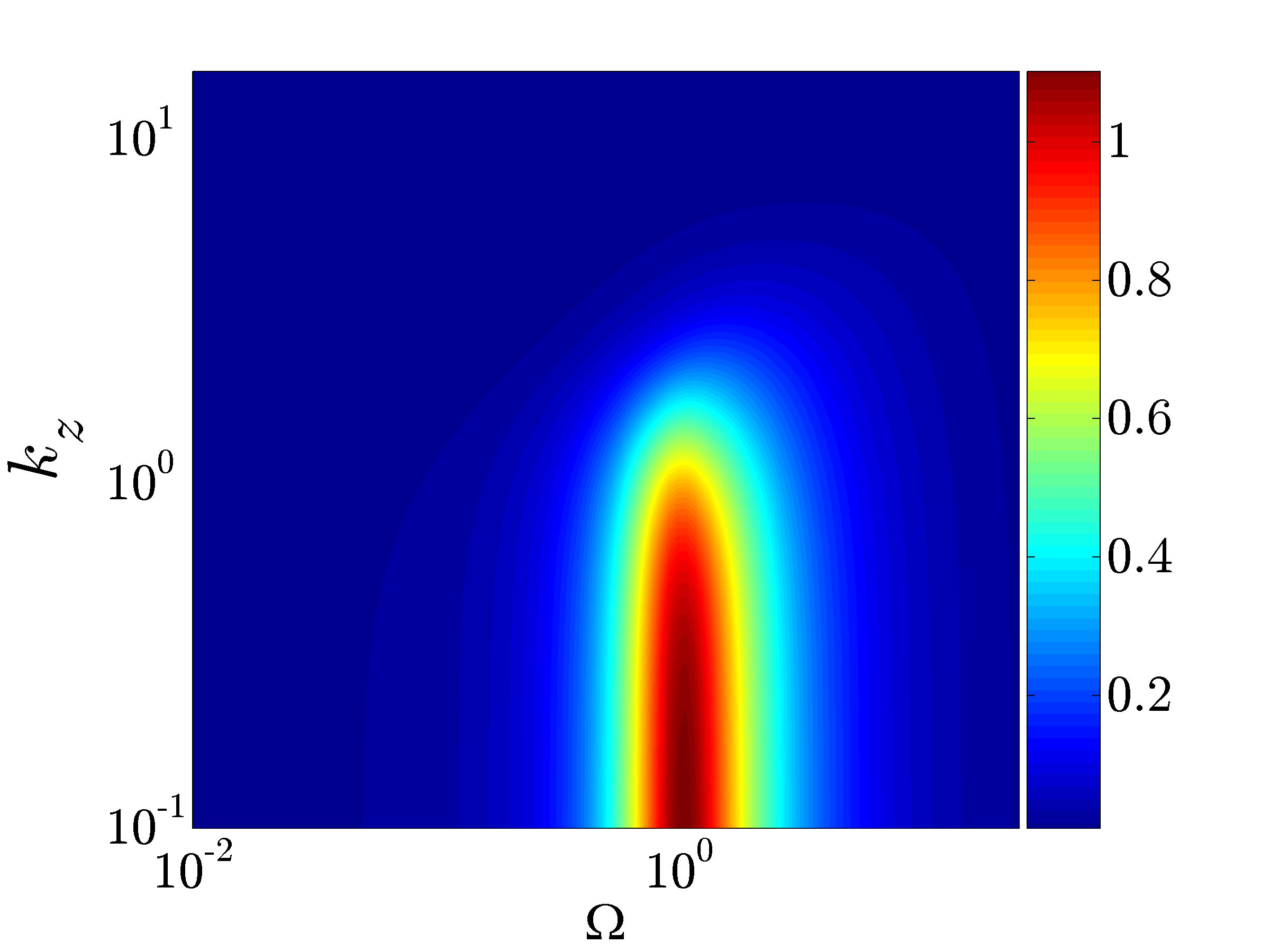}
    \end{tabular}
    \end{center}
    \caption{Plots of the base-flow-independent power spectral densities $\bPi_{rj}$ in Eq.~(\ref{eq.hs-component}) for $\beta \, = \, 0.1$, $\mu \, = \, 10$.}
    \label{Fig.Hrs}
    \end{figure}

\begin{figure}
    \begin{center}
    \begin{tabular}{cc}
    $\bPi_{u 2} (k_z,\Omega;\beta,\mu)$
    &
    $\bPi_{u 3} (k_z,\Omega;\beta,\mu)$
    \\
    \includegraphics[width=0.5\textwidth]{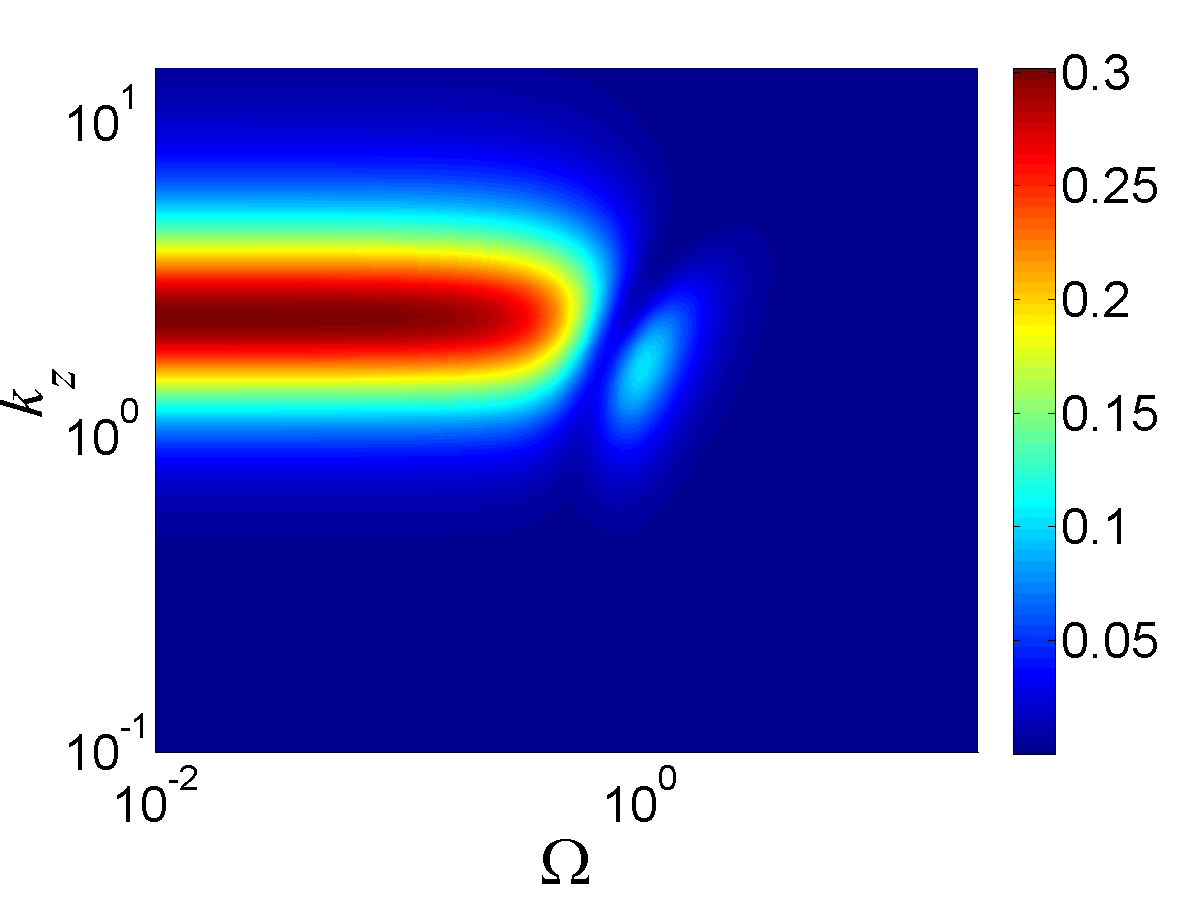}
    &
    \includegraphics[width=0.5\textwidth]{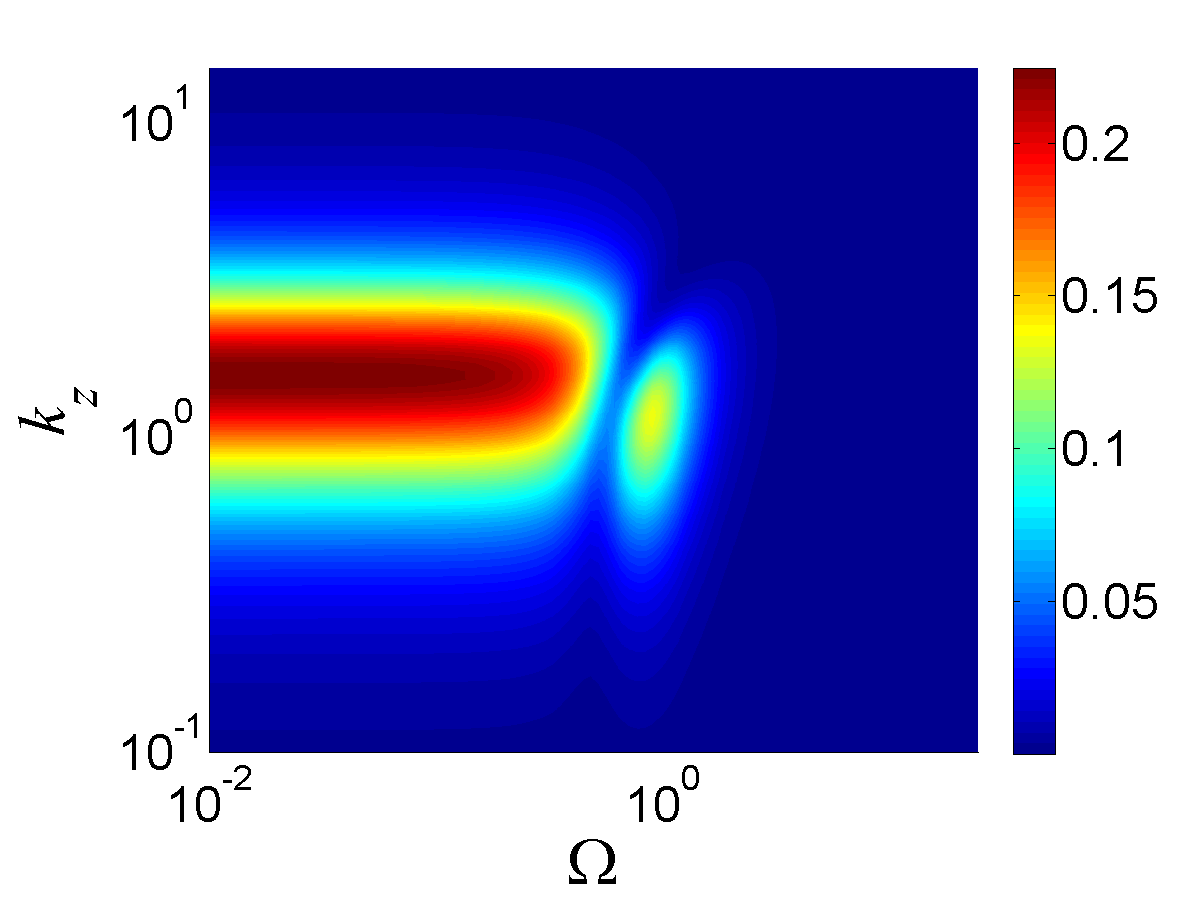}
    \\
    $\bPi_{u 2} (k_z,\Omega;\beta,\mu)$
    &
    $\bPi_{u 3} (k_z,\Omega;\beta,\mu)$
    \\
    \includegraphics[width=0.5\textwidth]{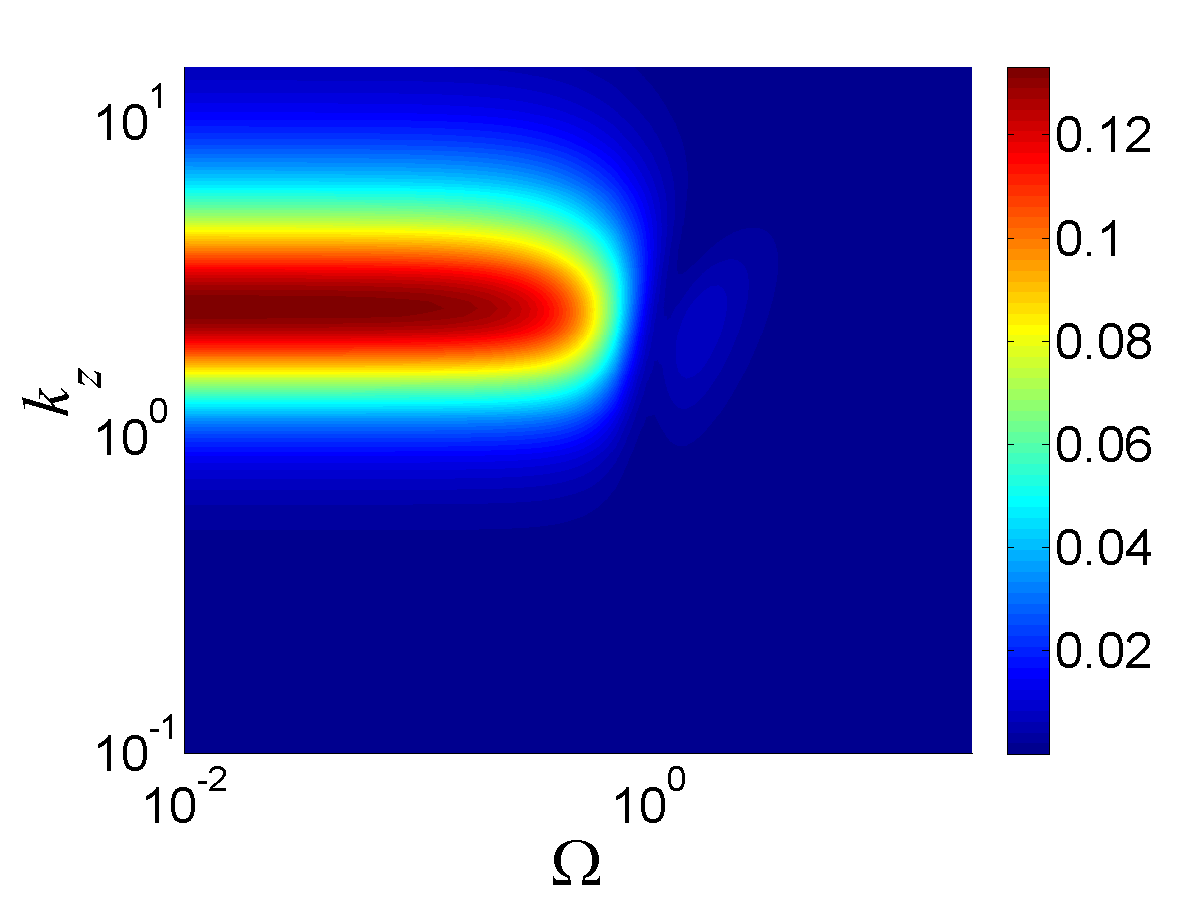}
    &
    \includegraphics[width=0.5\textwidth]{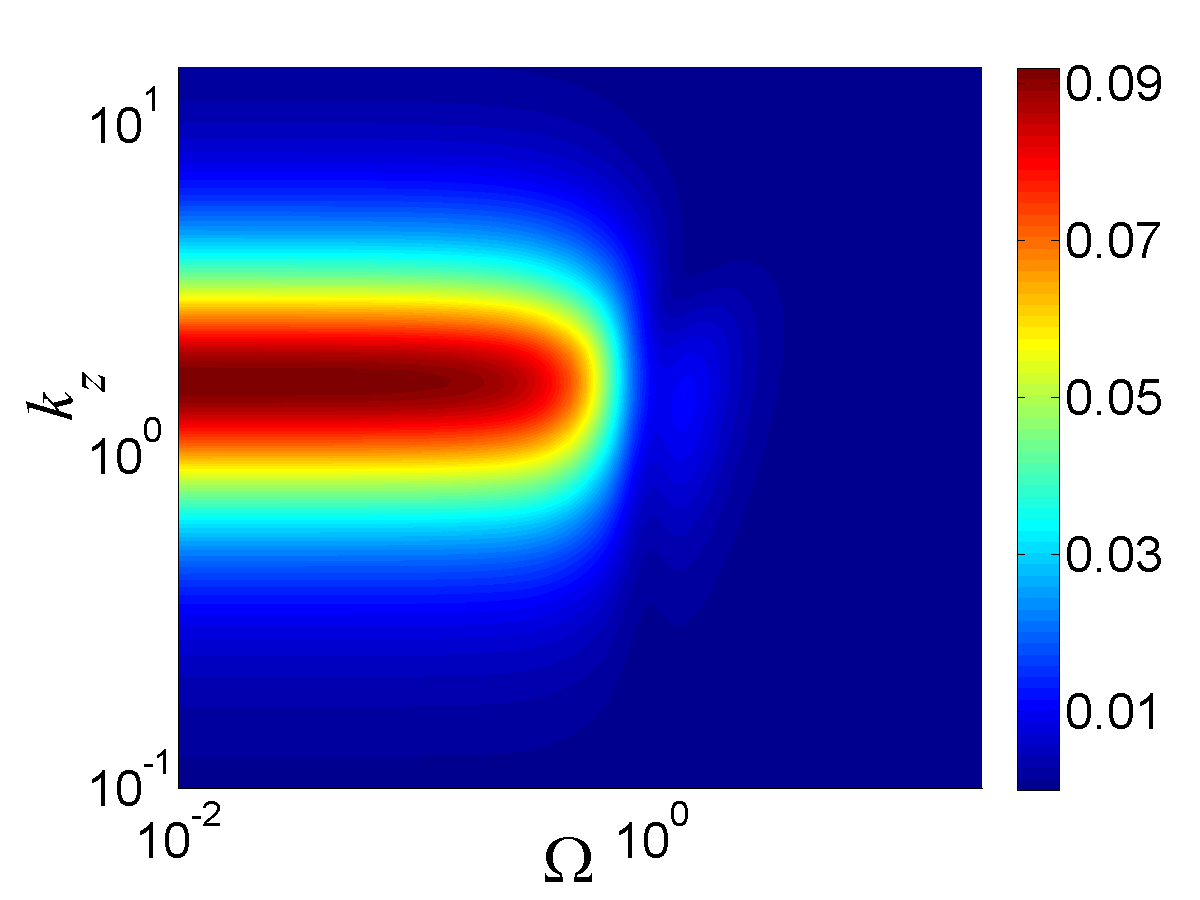}
    \end{tabular}
    \end{center}
    \caption{Plots of the base-flow-dependent power spectral densities $\bPi_{u2}$ and $\bPi_{u3}$ in Eq.~(\ref{eq.hs-component}) for Couette (first row) and Poiseuille (second row) flows with $\beta \, = \, 0.1$, $\mu \, = \,
    10$.}
    \label{Fig.Hus}
    \end{figure}

Figures~\ref{Fig.Hrs} and~\ref{Fig.Hus}, respectively, show the
($\Omega,k_z$)-dependence of the $Re$-independent power spectral densities $\bPi_{rj}$ in Eq.~(\ref{eq.hs-component}), for $\beta \, = \, 0.1$
and $\mu \, = \, 10$. As noted in \S~\ref{sec.frRe}, only $\bPi_{u2}$ and $\bPi_{u3}$ depend on the base velocity and polymer stresses. In view of this, the results in figure~\ref{Fig.Hus} are computed in both Couette and Poiseuille
flows. Since, at $k_x = 0$, $d_1$ does not affect $v$ and $w$, we do
not plot $\bPi_{v1}$  and $\bPi_{w1}$ in figure~\ref{Fig.Hrs}. Also, since
$\bPi_{w2} \, = \, \bPi_{v3}$, we only plot $\bPi_{v3}(k_z,\Omega;\beta,\mu)$.
We now discuss some important observations concerning the
results presented in these figures.

It is clearly seen that several frequency-response components peak at non-zero $\Omega$ values. This is in contrast to Newtonian fluids, where all power spectral
densities attain their respective maxima at $\Omega \, = \,
0$~\cite[][]{Mihailo2006}. Also, since the peaks for different
components of the frequency response are observed at different
locations in the $(\Omega,k_z)$-plane, these plots suggest distinct
amplification mechanisms. It is worth mentioning that the locations
of the peaks shift depending upon the $\beta$ and $\mu$ values. Our
results indicate that viscoelastic effects introduce additional
timescales which promote development of spatio-temporal flow
patterns with smaller time constants compared to Newtonian fluids.

At low $Re$ values $(\le 1)$, depending upon $\mu$, either
input-output amplification from $d_1$ to $u$ attains the largest
value or the amplification from ($d_2,d_3$) to $u$ attains the
largest value.  At small values of $\mu$, $\bPi_{u1}$ has the largest
magnitude; this suggests that at small Reynolds numbers and small
elasticity numbers, the streamwise forcing has the strongest
influence (on the velocity) and the most powerful impact of this
forcing is on the streamwise velocity component. At higher values of
$\mu$, $\bPi_{u2}$ and $\bPi_{u3}$ achieve the largest magnitudes; this suggests that at higher elasticity numbers, the spanwise and wall-normal forces have the strongest influence (on the velocity) and that the streamwise velocity component is most energetic.

\begin{figure}
    \begin{center}
    \begin{tabular}{cc}
    $\bPi_{u 1}(k_z,\Omega;\beta,\mu)$ & $\bPi_{v 2}(k_z,\Omega;\beta,\mu)$
    \\
    \includegraphics[width=0.45\textwidth]{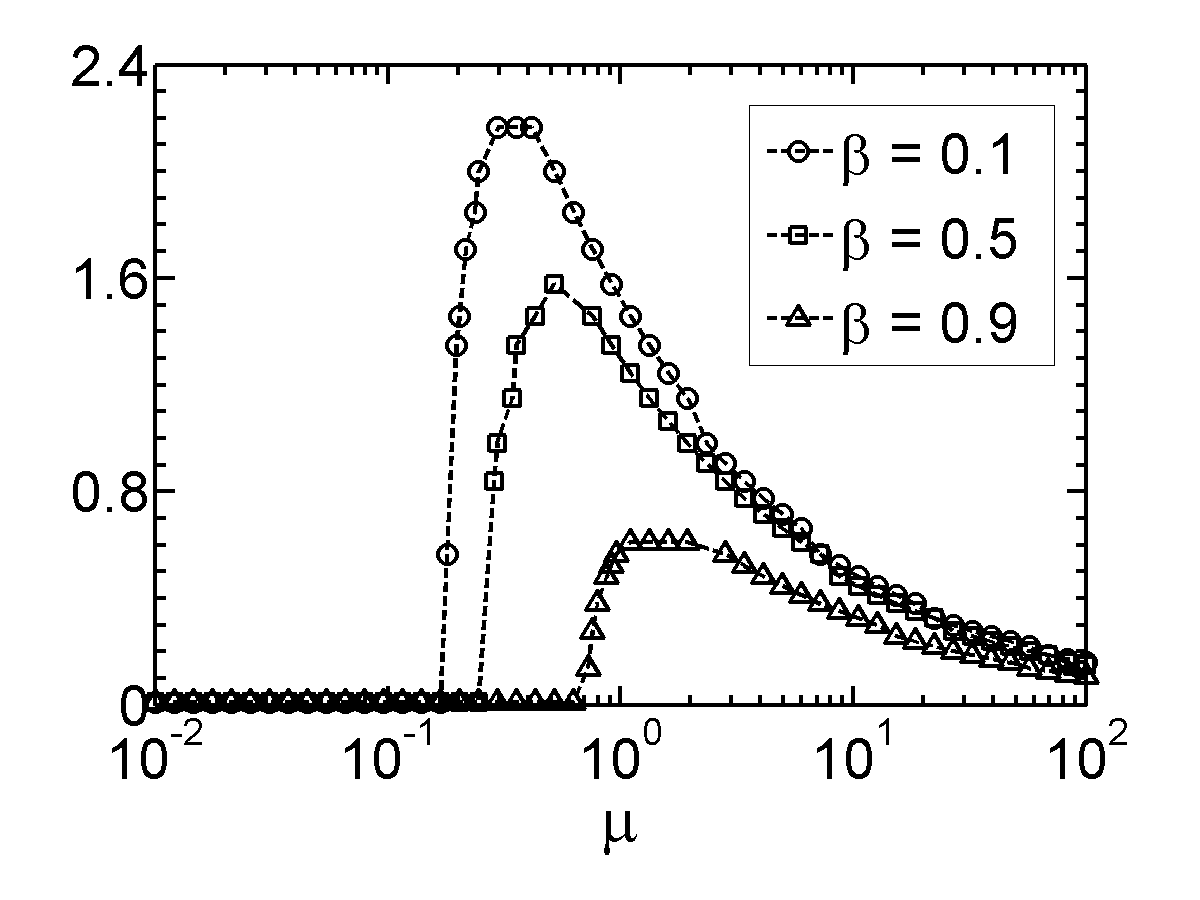}
    &
    \includegraphics[width=0.45\textwidth]{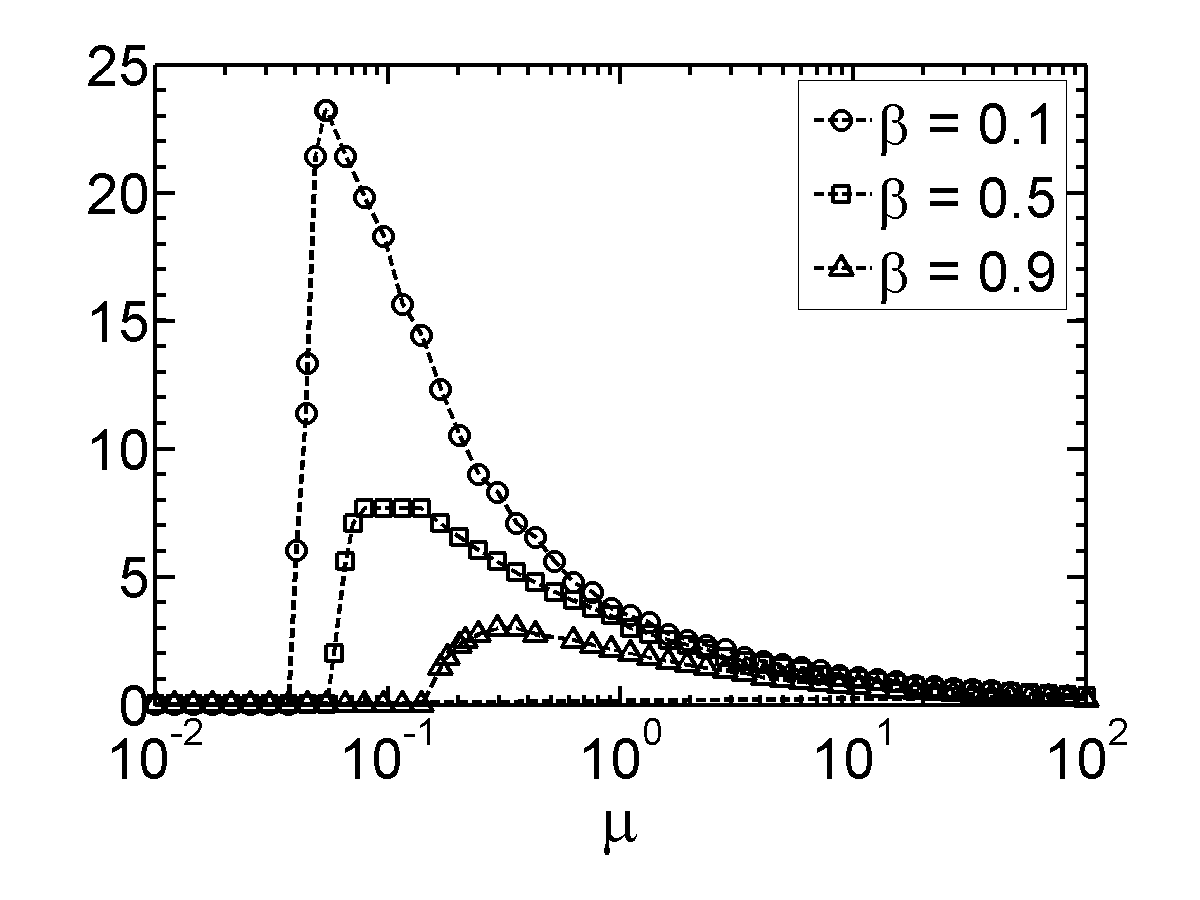}
    \\
    $\bPi_{v 3} (k_z,\Omega;\beta,\mu)$
    &
    $\bPi_{w 3} (k_z,\Omega;\beta,\mu)$
    \\
    \includegraphics[width=0.45\textwidth]{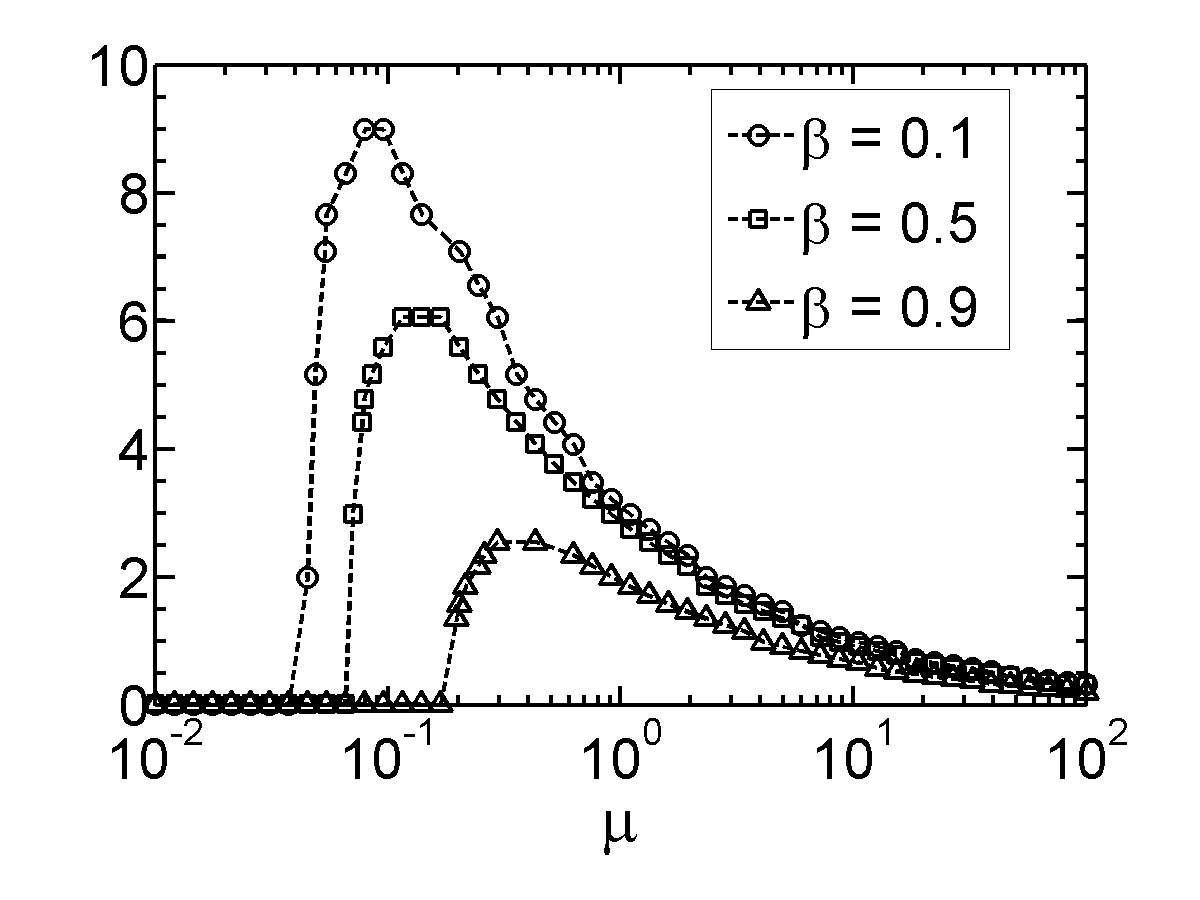}
    &
    \includegraphics[width=0.45\textwidth]{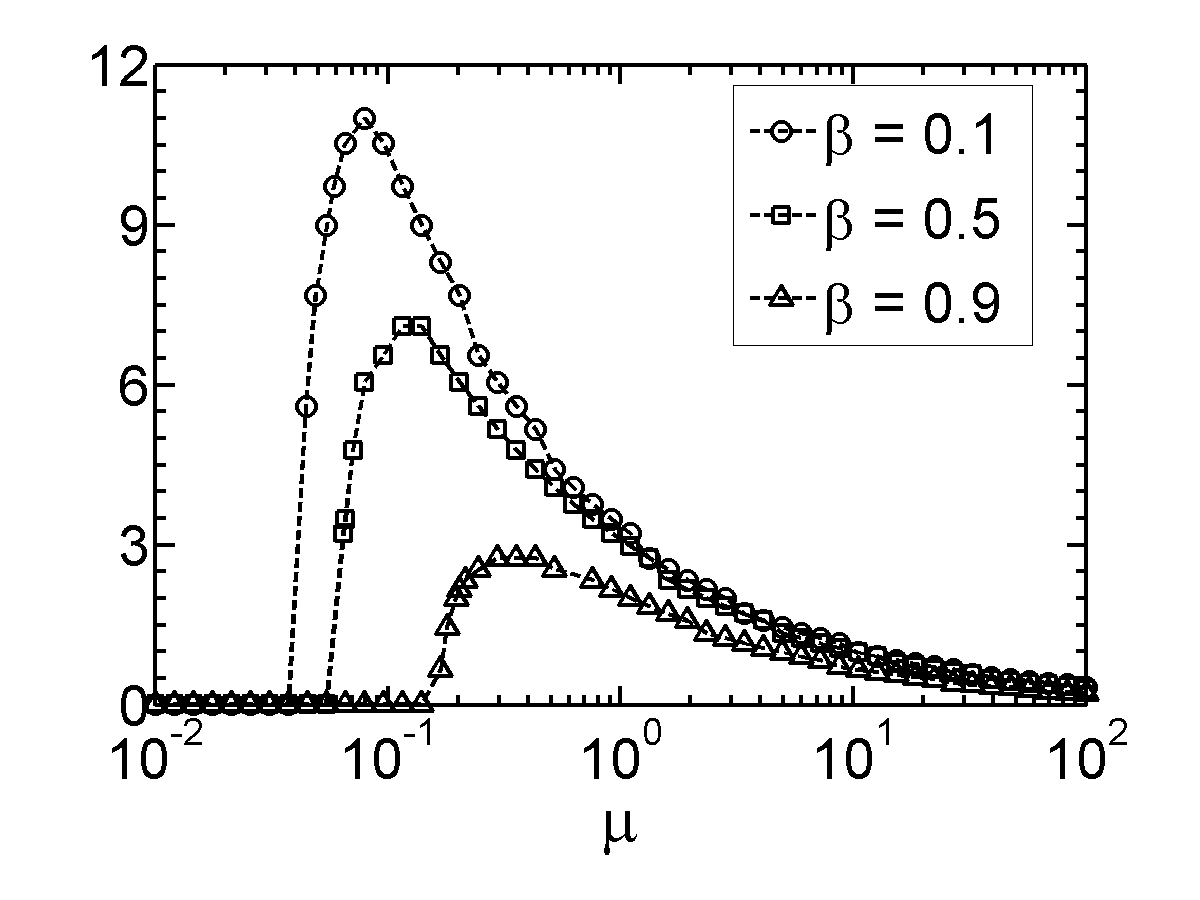}
    \end{tabular}
    \end{center}
    \caption{Variation in $\Omega_{\max}$ with $\mu$ for
    $\beta \, = \, \{0.1$, $0.5$, $0.9 \}$.}
    \label{Fig.Omegamax-Couette}
    \end{figure}

\begin{figure}
    \begin{center}
    \begin{tabular}{cc}
    $\bPi_{u 2} (k_z,\Omega;\beta,\mu)$
    &
    $\bPi_{u 3}(k_z,\Omega;\beta,\mu)$
    \\
    \includegraphics[width=0.45\textwidth]{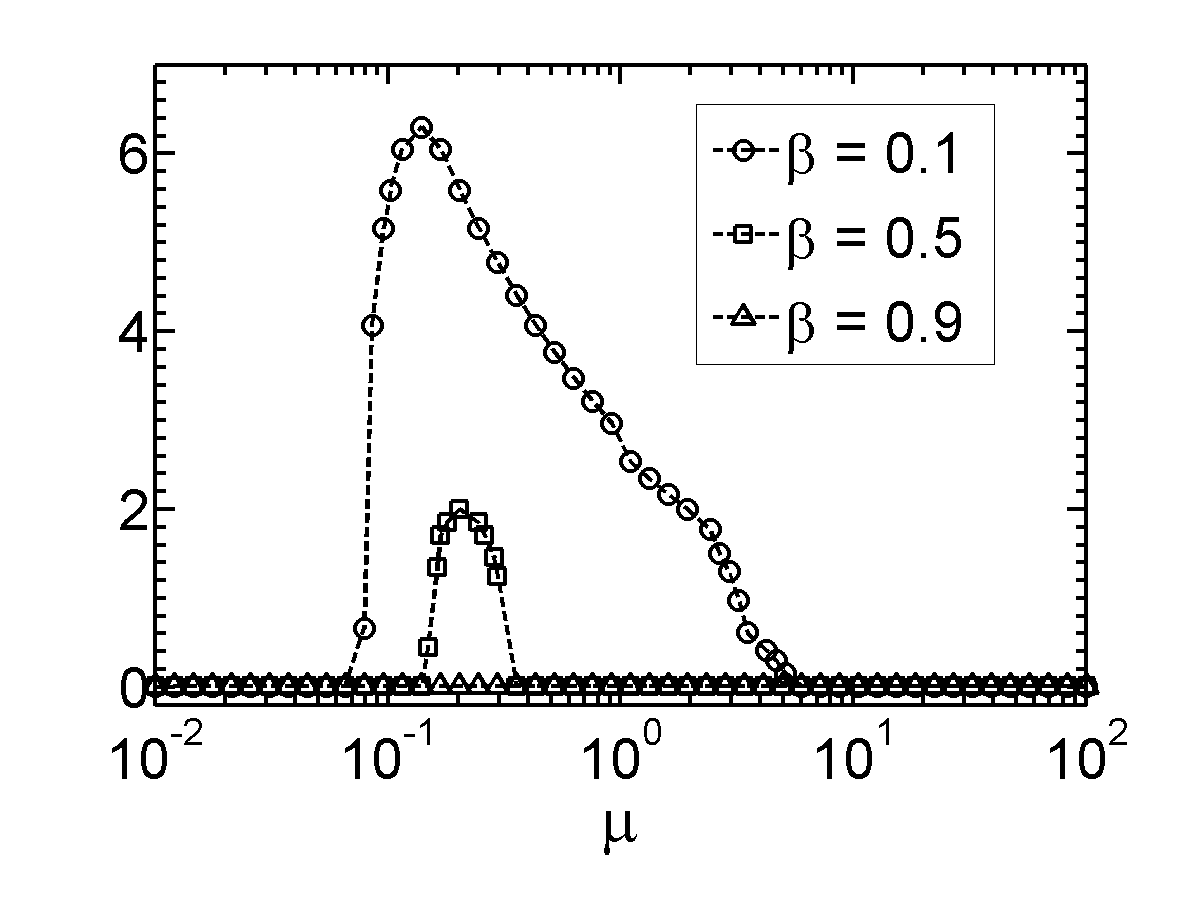}
    &
    \includegraphics[width=0.45\textwidth]{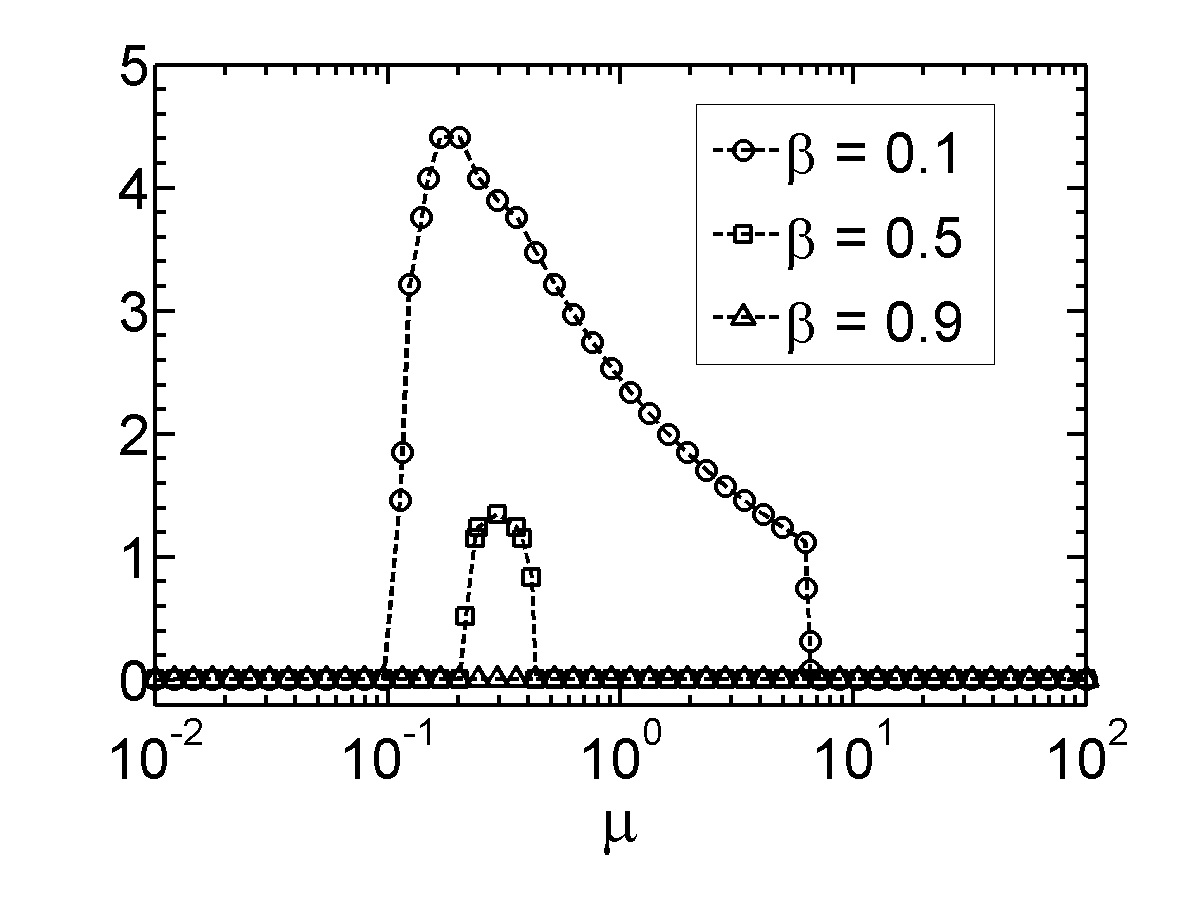}
    \\
    $\bPi_{u 2} (k_z,\Omega;\beta,\mu)$
    &
    $\bPi_{u 3} (k_z,\Omega;\beta,\mu)$
    \\
    \includegraphics[width=0.45\textwidth]{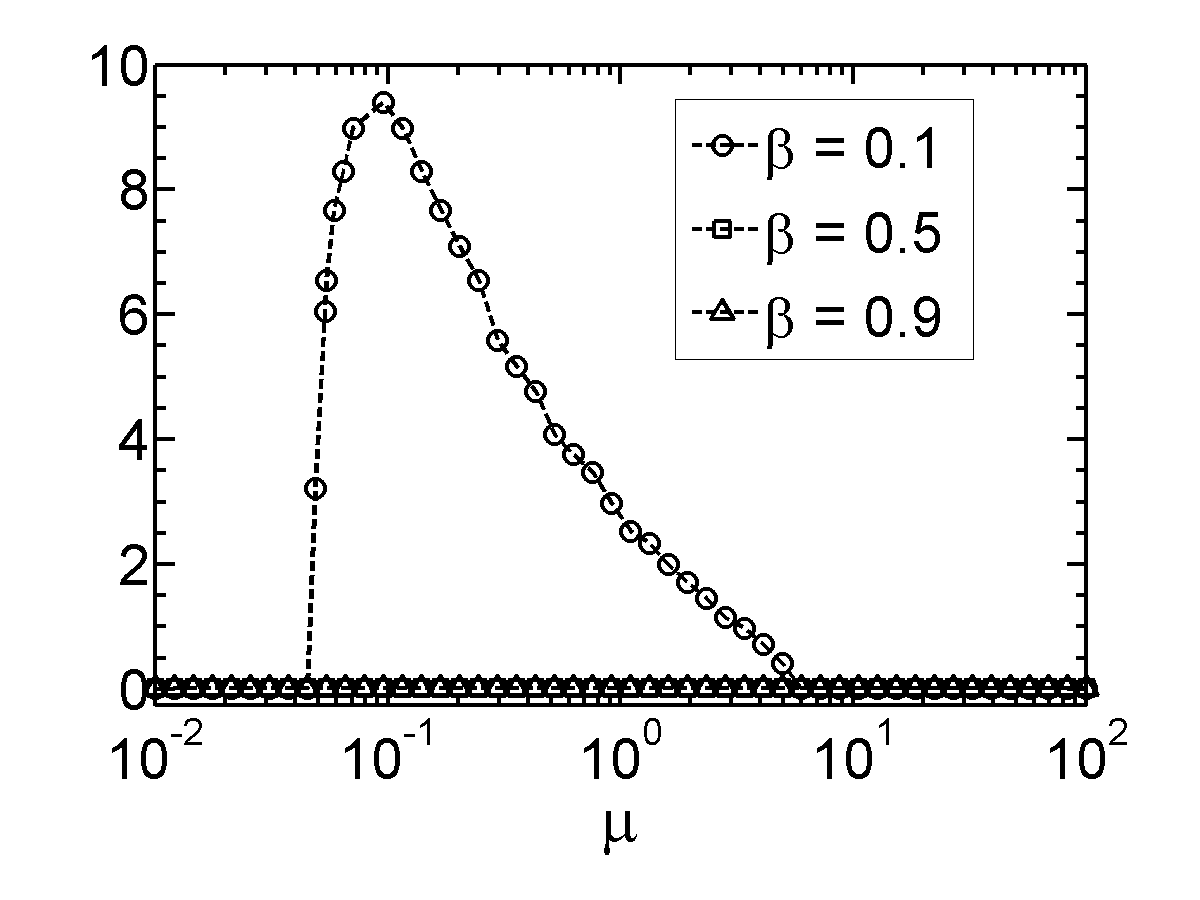}
    &
    \includegraphics[width=0.45\textwidth]{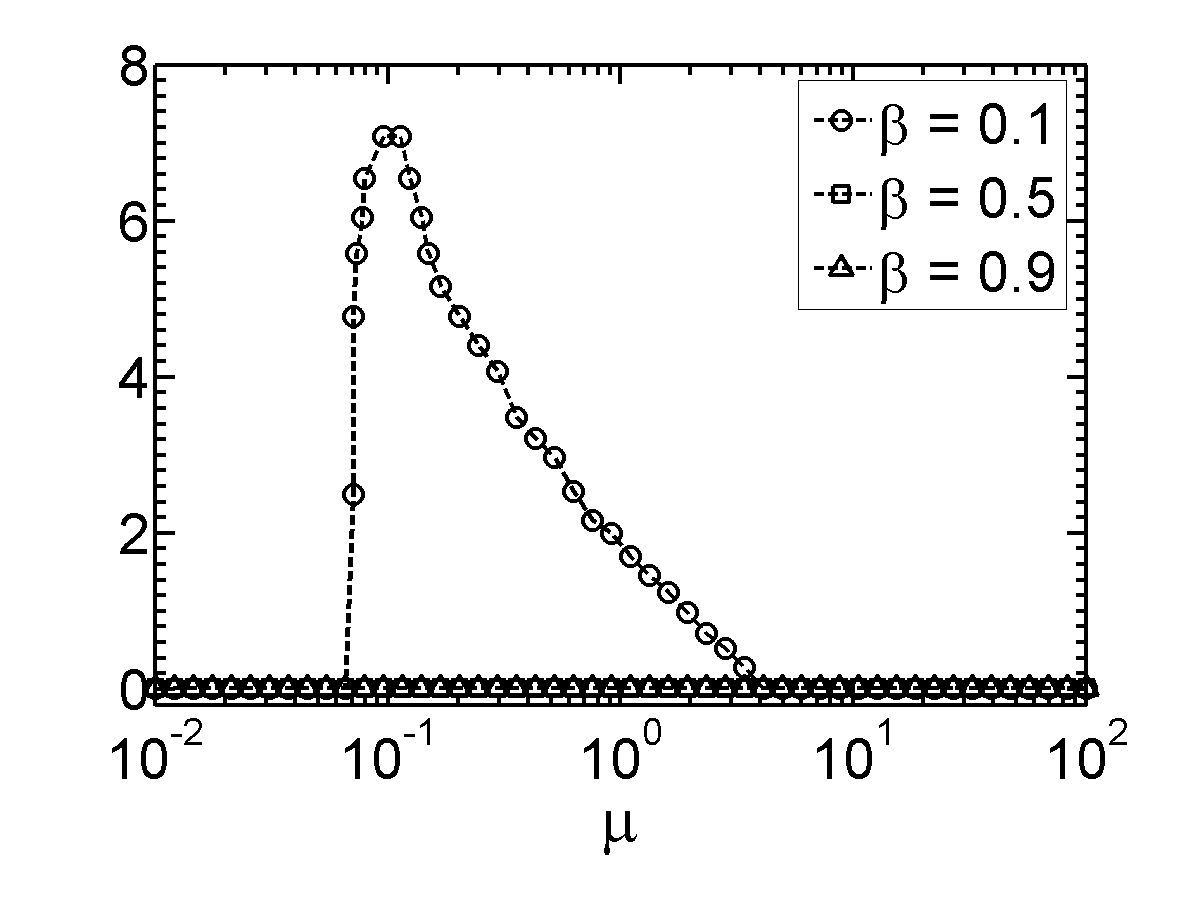}
    \end{tabular}
    \end{center}
    \caption{Variation in $\Omega_{\max}$ with $\mu$ for Couette (first row) and
    Poiseuille (second row) flows with $\beta \, = \, \{0.1$, $0.5$, $0.9 \}$.}
    \label{Fig.Omegamax-Poiseuille}
    \end{figure}

As discussed above, the power spectral density peaks (for different
components of the frequency response operator) are observed at
different locations in the $(\Omega,k_z)$-plane. We next analyze the
effects of parameters $\mu$ and $\beta$ on the magnitude of these
peaks and the locations of the respective maxima.
Figures~\ref{Fig.Omegamax-Couette}
and~\ref{Fig.Omegamax-Poiseuille}, respectively, illustrate the
variation with $\mu$ (for $\beta = \{ 0,1$, $0.5$, $0.9 \}$) in the
$\Omega$-value corresponding to the maxima of functions $\bPi_{rj}$ in Eq.~(\ref{eq.hs-component}); we denote this value by
$\Omega_{\max}$.  Below, we discuss the key features of these
results.

For $\mu$ greater than a certain threshold value, $\Omega_{\max}$
exhibits a maximum in $\mu$.  Also, the value of $\mu$
corresponding to the maximum in $\Omega_{\max}$ decreases with a
decrease in $\beta$. The above results suggest that for a particular
range of elasticity numbers, viscoelastic effects in Oldroyd-B
fluids promote amplification of flow structures with smaller time
constants than in Newtonian fluids. For small elasticity numbers,
figures~\ref{Fig.Omegamax-Couette} and~\ref{Fig.Omegamax-Poiseuille}
demonstrate that all power spectral densities achieve their peak
values at zero temporal frequency. This is consistent with the
behavior of Newtonian fluids~\cite[][]{Mihailo2006} as an Oldroyd-B
fluid is equivalent to a Newtonian fluid in the limit $\mu \,
\rightarrow \, 0$. The plots in figure~\ref{Fig.Omegamax-Couette}
also suggest that for large elasticity numbers, $\Omega_{\max}$
monotonically decreases (with a slow rate of decay) as $\mu$
increases.

It is also seen that $\Omega_{\max}$ increases with a decrease in $\beta$, suggesting the importance of effects with shorter time constants in viscoelastic fluids. In the limit $\beta \, \rightarrow \, 1$, $\Omega_{\max} \, = \, 0$ (results not shown), which is in agreement with the behavior of Newtonian
fluids~\cite[][]{Mihailo2006}; an Oldroyd-B fluid is equivalent to a
Newtonian fluid in the limit $\beta \, \rightarrow \, 1$.

For the base-flow-independent power spectral densities in
Eq.~(\ref{eq.hs-component}), a very good analytical estimate for
$\Omega_{\max}$ can be determined by projecting the operators in
$\bH_{rj}$ on the first eigenfunctions of $\bDelta^{-1} \bDelta^2$
(for $\bPi_{rj}$, $\{r=v,w; \; j=2,3\}$) and $\bDelta$ (for $\bPi_{u1}$).
(We refer the reader to Appendix~B of~\citet{Mihailo2005} for
spectral analysis of these two operators in the 2D/3C model.) Using this
approach, we determine the following expression for $\Omega_{\max}$
    \beq
    \label{eq.maxOmega}
    \Omega_{\max} \; = \;
    \left\{
    \ba{cl}
    \dfrac{\sqrt{{\sqrt{ \mu |\lambda_1 (k_z)| (1 - \beta) (\mu |\lambda_1 (k_z)| (1+\beta) + 2)}} \, - \, 1}}{\mu},
    & \text{$\mu \, > \, \dfrac{\sqrt{\frac{2}{1 \, - \, \beta}} \, - \,  1}{|\lambda_1 (k_z)|(1+\beta)}$},
    \\[0.25cm]
    0, & \text{otherwise,}
    \ea
    \right.
    \non
    \eeq
where $\lambda_1 (k_z)$ is the principal eigenvalue of the
underlying operator ($\bDelta$ for $\bPi_{u1}$; $\bDelta^{-1}\bDelta^2$
for $\bPi_{rj}$, $\{r=v,w; \; j=2,3\}$). From this expression, it
follows that $\Omega_{\max}$ increases with a decrease in $\beta$
and that it exhibits a maximum in $\mu$, explaining the trends
observed in figure~\ref{Fig.Omegamax-Couette}. Furthermore, for $\mu
\, \gg \, 1$, $\Omega_{\max}$ approximately scales as
$1/\sqrt{\mu}$, which justifies our earlier claim regarding the slow
rate of decay of $\Omega_{\max}$ with $\mu$ for large elasticity
numbers.

\begin{figure}
    \begin{center}
    \begin{tabular}{cc}
    $\mu \, = \, 0.1$ & $\mu \, = \, 2$
    \\
    \includegraphics[width=0.5\textwidth]{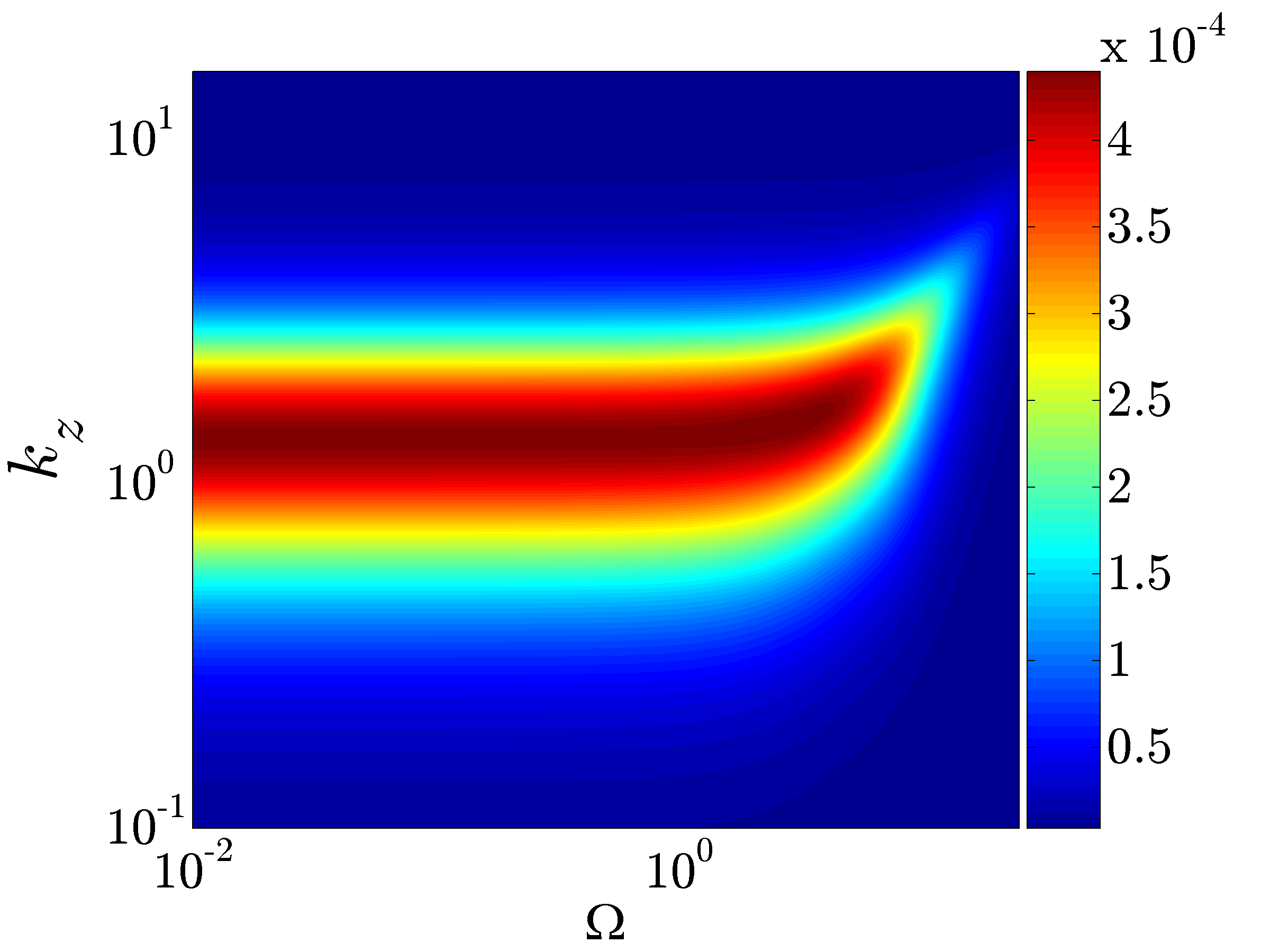}
    &
    \includegraphics[width=0.5\textwidth]{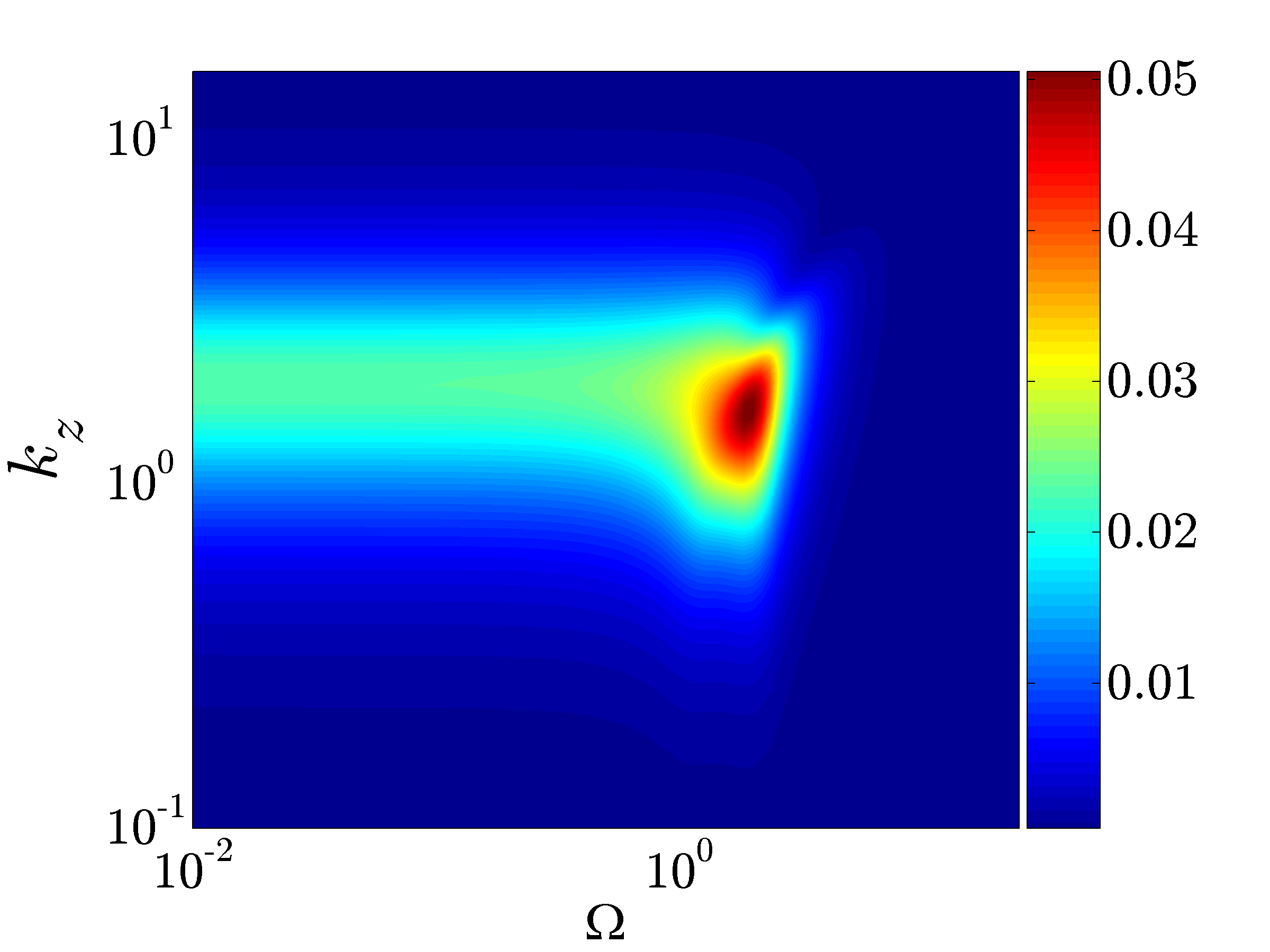}
    \\
    $\mu \, = \, 10$ & $\mu \, = \, 100$
    \\
    \includegraphics[width=0.5\textwidth]{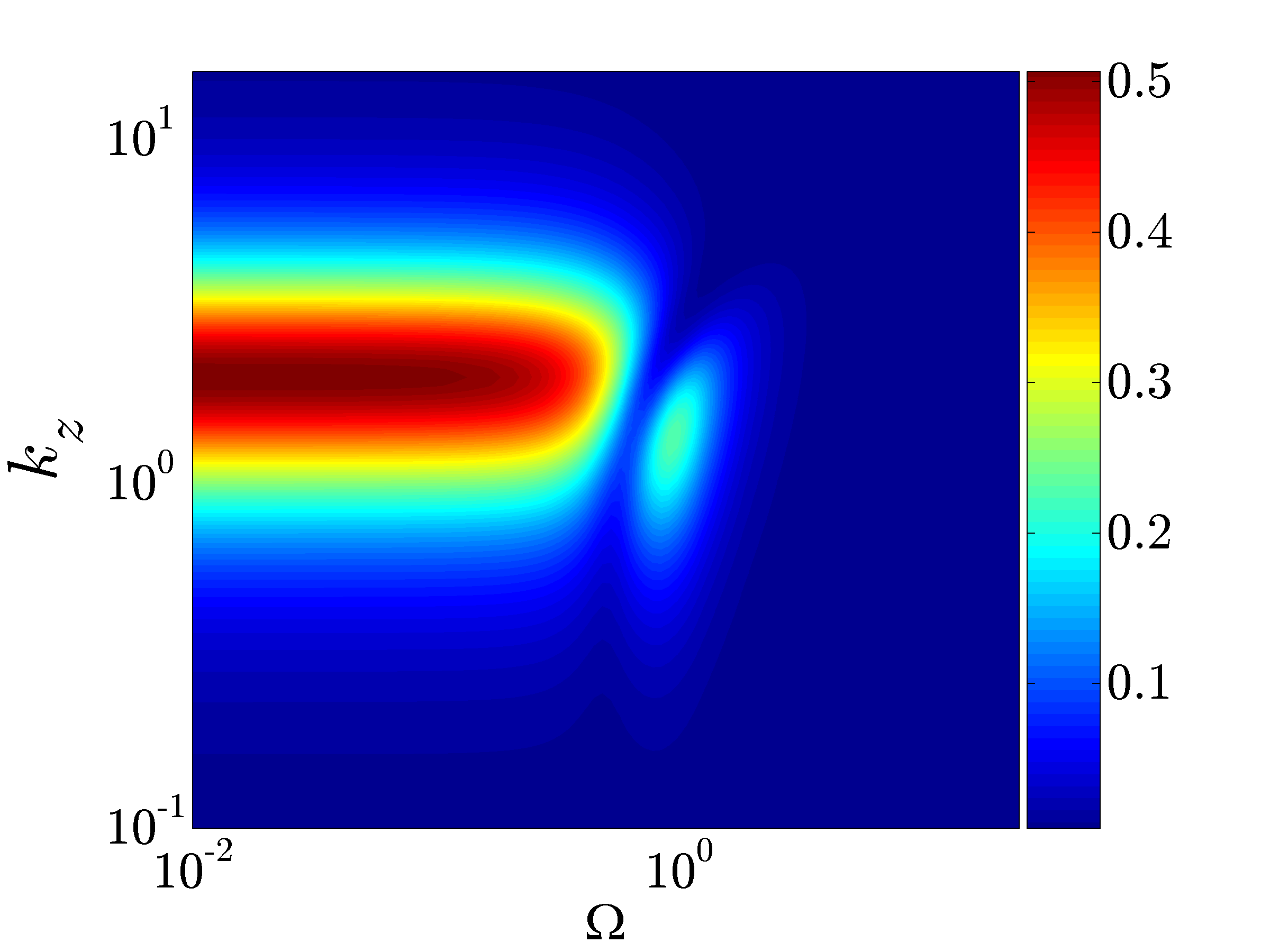}
    &
    \includegraphics[width=0.5\textwidth]{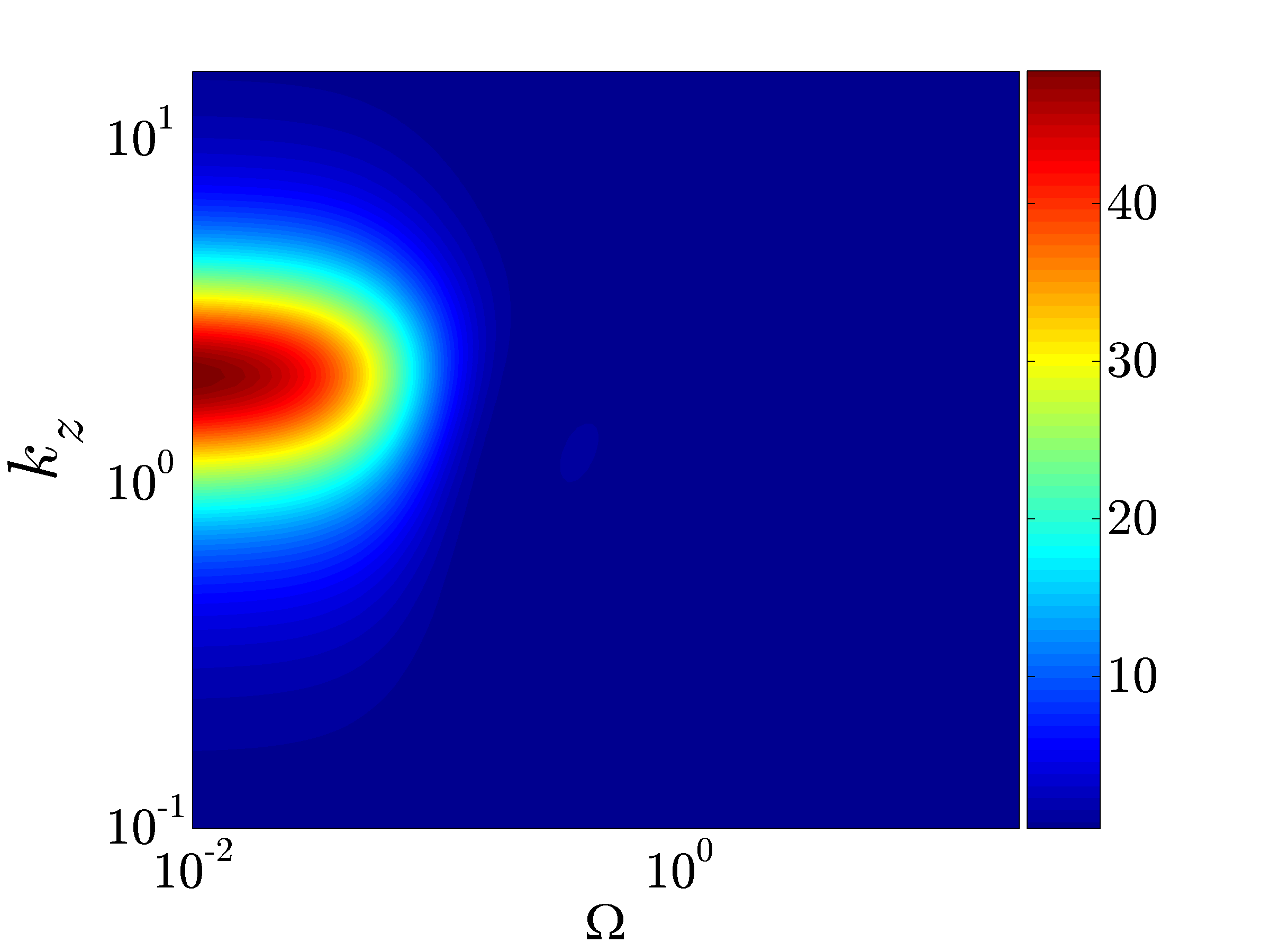}
    \end{tabular}
    \end{center}
    \caption{Plots of $\bPi_b(k_z,\Omega;\beta,\mu) \, = \,
    \bPi_{u 2}(k_z,\Omega;\beta,\mu) \, + \, \bPi_{u 3}(k_z,\Omega;\beta,\mu)$
    in Couette flow with $\beta \, = \, 0.1$, $\mu \, = \, \{0.1$, $2$, $10$, $100 \}$.}
    \label{Fig.HS-u-dv-Cou}
    \end{figure}

The analytical expressions for $\Omega_{\max}$ are much more
difficult to obtain for the base-flow-dependent power spectral densities $\bPi_{u2}$ and $\bPi_{u3}$. In spite of this, essential trends can be ascertained from figure~\ref{Fig.HS-u-dv-Cou}, which shows the plots of function $\bPi_b(k_z,\Omega;\beta,\mu)$ in Couette flow with $\beta \, = \,0.1$ and $\mu \, = \, \{0.1$, $2$, $10$, $100 \}$. (This
function quantifies the power spectral density of the frequency
response operator that maps ($d_2,d_3$) to streamwise velocity $u$
at $Re \, = \, 1$.) For $\mu \, = \, 0.1$, the frequency response
achieves a global maximum at $\Omega_{\max} \, = \, 0$, but the broad
spectrum in $\Omega$ around $k_z \, \approx \, O(1)$ indicates that
the large values are maintained up until $\Omega \, \approx \, 5$.
On the other hand, at $\mu \, = \, 2$, the global peak is located in
the narrow region around $\Omega_{\max} \, \approx \, 2$. With a
further increase in elasticity number, two competing peaks at zero
and $O(1)$ temporal frequencies appear; finally, for large values of
$\mu$ the spectrum peak shifts to the narrow region around zero
temporal frequency, with $\Omega_{\max} \, = \, 0$.

We have also studied the effects of $\mu$ and $\beta$ on the maximum
value of power spectral densities. We briefly highlight several
important points. First,
the maximum values of all power spectral densities decrease with an increase in $\beta$ and a decrease in $\mu$.  This suggests that amplification becomes weaker as one approaches the Newtonian fluid limit. A similar dependence of the maximum value of the growth function on Deborah (Weissenberg) number was reported by~\citet{Sureshkumar1999} in their study of two-dimensional time-dependent simulations of creeping plane Couette flow of Oldroyd-B fluids.
Second, the peak values of the base-flow-dependent power spectral densities $\bPi_{u2}$ and $\bPi_{u3}$ monotonically increase with $\mu$. On the other hand, the peak values of the base-flow-independent power spectral densities $\bPi_{u1}$ and $\bPi_{rj}$, $\{r=v,w; \; j=2,3\}$, first increase with an increase in $\mu$ and then plateau after $\mu$ becomes sufficiently large. The monotonic increase of $\bPi_{u2}$ and $\bPi_{u3}$  with the elasticity number demonstrates the significance of the coupling operator $\bC_{p2}$, which captures the work done by the polymer stresses on the flow. An in-depth study of the physical mechanisms behind this viscoelastic amplification is given in \S~\ref{sec.Energy}.
Third, the spanwise wavenumbers corresponding to the maxima in the components of the frequency response increase with an increase in $\mu$ and a decrease in $\beta$.  This suggests that the dominant structures become less spread in the spanwise direction with an increase in $\mu$ and a decrease in $\beta$.

\section{Energy amplification mechanisms}
    \label{sec.Energy}

In order to elucidate the energy amplification mechanisms in
Oldroyd-B fluids, we next analyze the Reynolds-Orr equation for
streamwise-constant channel flow. As is well known, the Reynolds-Orr
equation describes the evolution of the energy of velocity
fluctuations around a given base flow condition~\citep{schhen01}. In
our study, the initial conditions on velocity and polymer stress
fluctuations are set to zero, but the flow is driven by the
spatio-temporal stochastic body forcing $\bd$. This random body
forcing generates the velocity field $\bv$ and polymer stresses
$\btau$, which are also of stochastic nature~\cite[][]{Farrell1993}.

The energy-evolution equation is derived by multiplying the
Navier-Stokes equations by the velocity vector, followed by
integration over the wall-normal direction and ensemble averaging in
time. The equations are further simplified using the divergence
theorem and the boundary conditions on $\bv$. For
streamwise-constant perturbations, the Reynolds-Orr equation for an
Oldroyd-B fluid is given by:
    \beq
    \ba{l}
    \dfrac{1}{2}
    \dfrac{\mrd E}{\mrd t}
    \; = \;
    -\, \dfrac{\beta k_z^2}{Re} E
    \; - \;
    \inprod{u}{U' v}
    \; + \;
    \inprod{\bv}{\bd}
    \; + \;
    \dfrac{\beta}{Re}
    \left(
    \inprod{u}{\pyy u}
    \, + \,
    \inprod{v}{\pyy v}
    \, + \,
    \inprod{w}{\pyy w}
    \right)
    \\[0.3cm]
    -\, \dfrac{1 \, - \, \beta}{Re}
    \left(
    \inprod{\tau_{xy}}{\py u}
    \, + \,
    \inprod{\tau_{yz}}{\py v}
    \, + \,
    \inprod{\tau_{xz}}{\py w}
    \, + \,
    \mri k_z
    \left(
    \inprod{\tau_{xz}}{u}
    \, + \,
    \inprod{\tau_{yz}}{v}
    \, + \,
    \inprod{\tau_{zz}}{w}
    \right)
    \right),
    \ea
    \non
    \eeq
where $\inprod{\cdot}{\cdot}$ denotes integration in $y$ and
ensemble averaging in $t$, and $E \, = \, E(t,k_z;Re,\beta,\mu)$
represents the kinetic energy, that is
    \beq
    E
    \; = \;
    \inprod{\bv}{\bv}
    \; = \;
    \int_{-1}^{1}
    {\cal E}
    \left( \bv^{*}(y,k_z,t) \bv(y,k_z,t) \right) \, \mrd y.
    \non
    \eeq
Here, the asterisk denotes the complex-conjugate-transpose of vector $\bv$,
and ${\cal E}$ denotes ensemble averaging~\cite[][]{mccomb91}
    \beq
    {\cal E}
    \left( \bv (\cdot,t) \right)
    ~=~
    \lim_{T \,\rightarrow\, \infty}
    \dfrac{1}{T}
    \int_{0}^{T} \bv(\cdot, t + \tau) \, \mrd \tau.
    \non
    \eeq
In the steady-state limit, the kinetic energy $E$ is given by
    \beq
    \ba{l}
    E
    \; = \;
    \dfrac{Re}{\beta k_z^2}
    \left(
    -\langle u, U' v \rangle
    \; + \;
    \inprod{\bv}{\bd}
    \right)
    \; + \;
    \dfrac{1}{k_z^2}
    \left(
    \inprod{u}{\pyy u}
    \, + \,
    \inprod{v}{\pyy v}
    \, + \,
    \inprod{w}{\pyy w}
    \right)
    \\[0.3cm]
    - \,
    \dfrac{1\,-\,\beta}{\beta k_z^2}
    \left(
    \inprod{\tau_{xy}}{\py u}
    \, + \,
    \inprod{\tau_{yz}}{\py v}
    \, + \,
    \inprod{\tau_{xz}}{\py w}
    \, + \,
    \mri k_z
    \left(
    \inprod{\tau_{xz}}{u}
    \, + \,
    \inprod{\tau_{yz}}{v}
    \, + \,
    \inprod{\tau_{zz}}{w}
    \right)
    \right).
    \ea
    \non
    \eeq
In this equation, the first two terms on the right-hand-side are
contributions due to the Reynolds stress and the work done by the
body forces; the third group of terms accounts for viscous
dissipation, and the last group of terms corresponds to the work
done by the polymer stresses on the flow. This last contribution to kinetic
energy is -- in general -- sign indefinite and it is also referred
to as an energy-exchange term~\cite[][]{Schumacher2006}.

Two points are worth highlighting.  First, as in Newtonian fluids~\cite[][]{schhen01}, the nonlinear terms do not contribute to the kinetic energy (they are conservative and only redistribute energy between different modes). Thus, the Reynolds-Orr
equations for viscoelastic fluids derived using linearized and fully
non-linear equations correspond to each other.  Second, the steady-state energy $E$ defined here is exactly the same as the ensemble-average energy density defined in~\cite{Nazish2008a}. As a matter of fact, this quantity is precisely determined by the $H_2$ norm of a stochastically forced linearized system. Thus, the
energy density of the LNSE can be efficiently computed using Lyapunov equations, which circumvents the need for running costly stochastic numerical simulations.

The explicit scaling of the steady-state energy density with $Re$ is
given in \S~\ref{sec.frRe}. Based on this, for
asymptotically large times, $E$ can be written as:
    \begin{equation}
    E(k_z;Re,\beta,\mu)
    \; = \;
    f (k_z;\beta,\mu)
    Re
    \; + \;
    g (k_z;\beta,\mu)
    Re^3,
    \non
    \end{equation}
where functions $f$ and $g$ correspond to the Reynolds-number-independent terms in Eq.~(\ref{eq.E-total}). This equation suggests that at higher $Re$ values, $g$ is expected to contribute most to the energy, whereas at smaller $Re$ values, $f$ is expected to contribute most to the energy; we will show that the latter
observation holds only at moderate values of $\mu$ and $\beta$. From
\S~\ref{sec.frRe}, it follows that only streamwise
velocity contributes to $g$; this contribution arises due to
amplification from $d_2$ and $d_3$ to $u$. On the other hand,
amplification from $d_1$ to $u$ and $(d_2,d_3)$ to $(v,w)$ is
captured by function $f$. The explicit $Re$-scaling of various terms
in the steady-state Reynolds-Orr equation can be obtained by
analyzing the Lyapunov equation (see Appendix~\ref{sec.lyap}); let
the overbar designate the velocity and polymer stress fluctuations
at $Re = 1$, and let $u_1$ and $u_{2,3}$ denote the components of
$u$ arising due to the action of $d_1$ and ($d_2,d_3$),
respectively. Then, the $Re$-independent terms
    $
    f(k_z;\beta,\mu)
    $
and
    $
    g(k_z;\beta,\mu)
    $
in the expression for the steady-state energy are given by
    \beq
    \ba{rcl}
    f
    & \!\! = \!\! &
    \dfrac{1}{\beta k_z^2}
    \bigg(
    \inprod{\bar{\bv}}{\bd}
    \, + \,
    \beta
    \left(
    \inprod{\bar{u}_1}{\pyy \bar{u}_1}
    \, + \,
    \inprod{\bar{v}}{\pyy \bar{v}}
    \, + \,
    \inprod{\bar{w}}{\pyy \bar{w}}
    \right)
    \, - \,
    (1-\beta)
    \big(
    \inprod{\bar{\tau}_{xy}}{\py \bar{u}_1}
    \\[0.25cm]
    & \!\! \!\! &
    + \,
    \inprod{\bar{\tau}_{yz}}{\py \bar{v}}
    \, + \,
    \inprod{\bar{\tau}_{xz}}{\py \bar{w}}
    \, + \,
    \mri k_z
    \left(
    \inprod{\bar{\tau}_{xz}}{\bar{u}_1}
    \, + \,
    \inprod{\bar{\tau}_{yz}}{\bar{v}}
    \, + \,
    \inprod{\bar{\tau}_{zz}}{\bar{w}}
    \right)
    \big)
    \bigg),
    \\[0.25cm]
    g
    & \!\! = \!\! &
    \dfrac{1}{\beta k_z^2}
    \left(
    -\inprod{\bar{u}}{U' \bar{v}}
    \, + \,
    \beta
    \inprod{\bar{u}_{2,3}}{\pyy \bar{u}_{2,3}}
    \, - \,
    (1 \, - \, \beta)
    \left(
    \inprod{\bar{\tau}_{xy}}{\py \bar{u}_{2,3}}
    \, + \,
    \mri k_z
    \inprod{\bar{\tau}_{xz}}{\bar{u}_{2,3}}
    \right)
    \right)
    \\[0.25cm]
    & \!\! = \!\! &
    g_{\inprod{u}{U' v}}
    \, + \,
    g_{\inprod{u}{\pyy u}}
    \, + \,
    g_{\inprod{\tau_{xy}}{\py u}}
    \, + \,
    g_{\inprod{\tau_{xz}}{\mri k_z u}}.
    \ea
    \non
    \eeq

\begin{figure}
    \begin{center}
    \begin{tabular}{cc}
    $f(k_z;\beta,\mu)$
    &
    $g(k_z;\beta,\mu)$
    \\
    \includegraphics[width=0.5\textwidth]{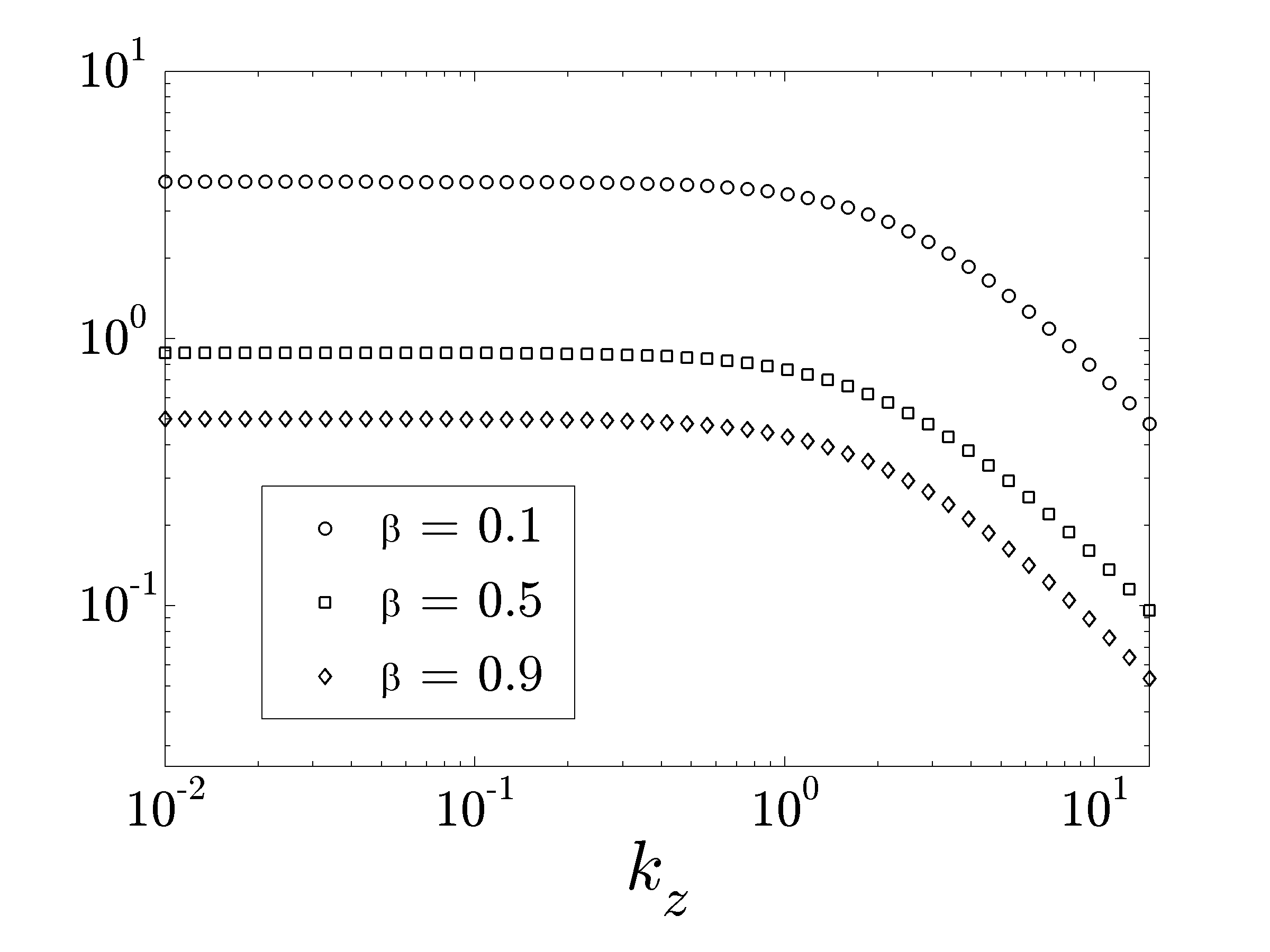}
    &
    \includegraphics[width=0.5\textwidth]{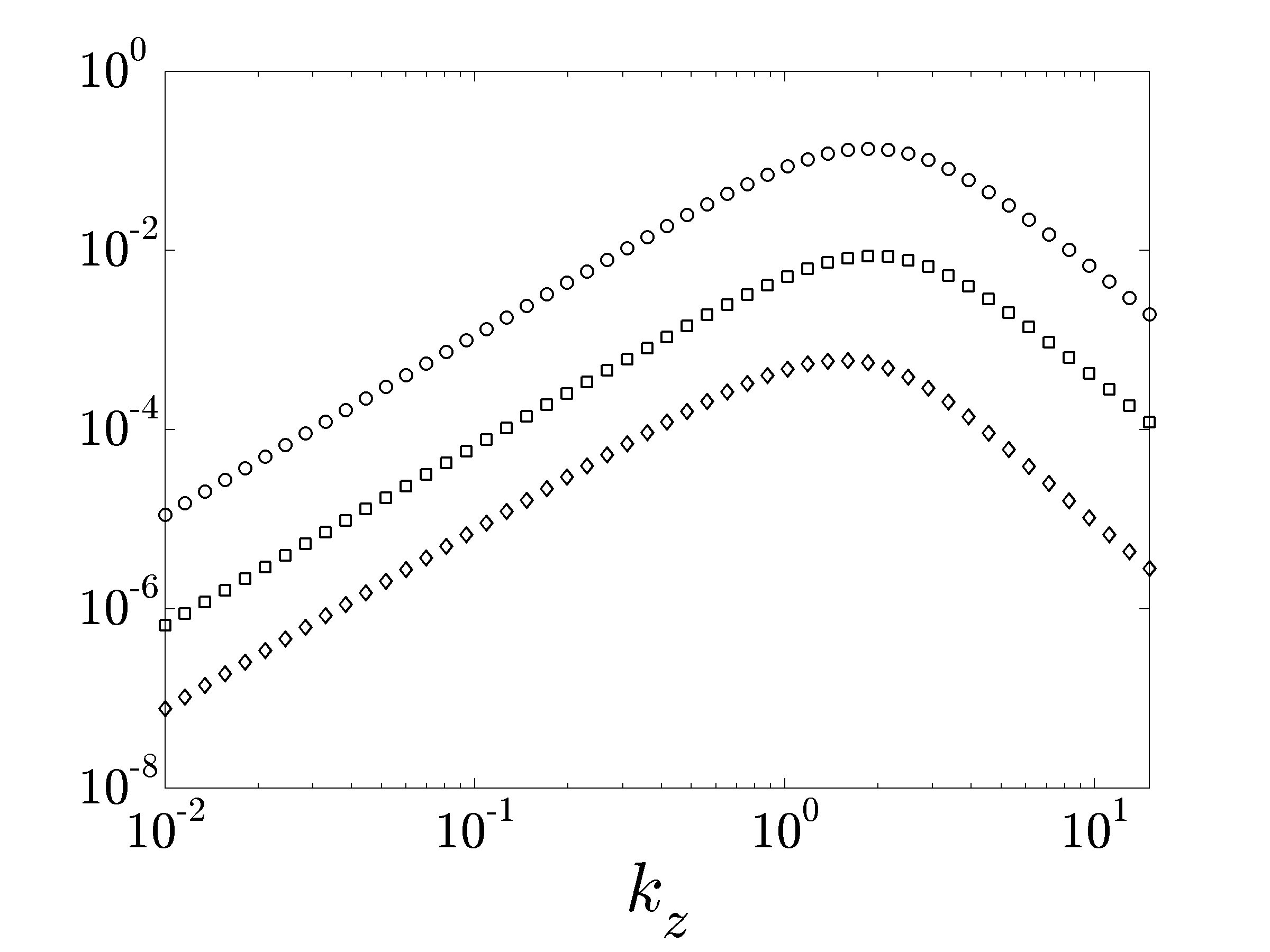}
    \end{tabular}
    \end{center}
    \caption{Variation in $f$ and $g$ with $k_z$ for $\mu \, = \, 10$ and
    $\beta \, = \, \{0.1$, $0.5$, $0.9 \}$; $g$ in Couette flow is
    shown.} \label{Fig:ERebeta}
    \end{figure}

\begin{figure}
    \begin{center}
    \begin{tabular}{ccc}
    $f(k_z;\beta,\mu)$
    &
    $g(k_z;\beta,\mu)$
    &
    $g(k_z;\beta,\mu)/\mu$
    \\
    \includegraphics[width=0.35\textwidth]{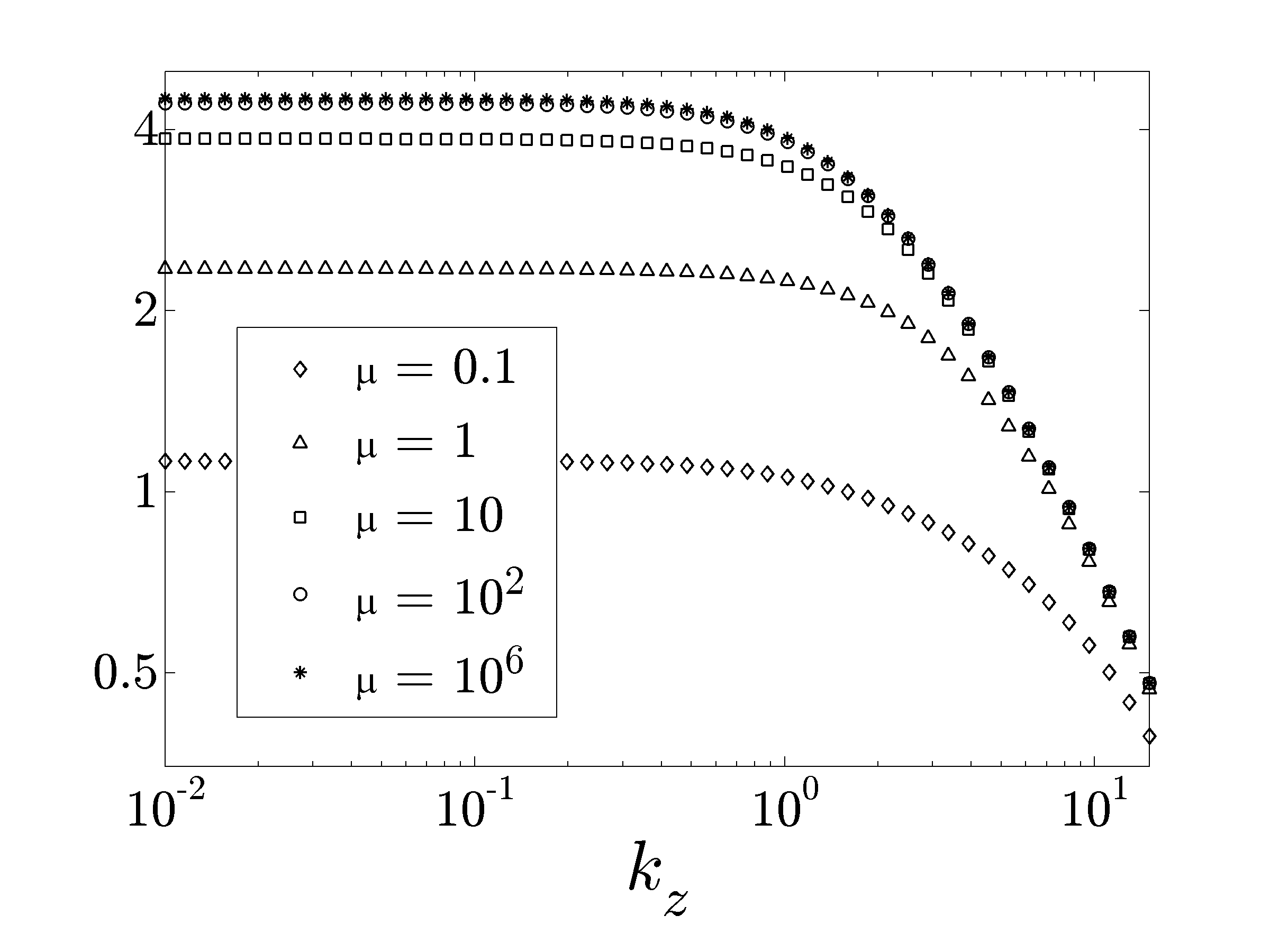}
    &
    \includegraphics[width=0.35\textwidth]{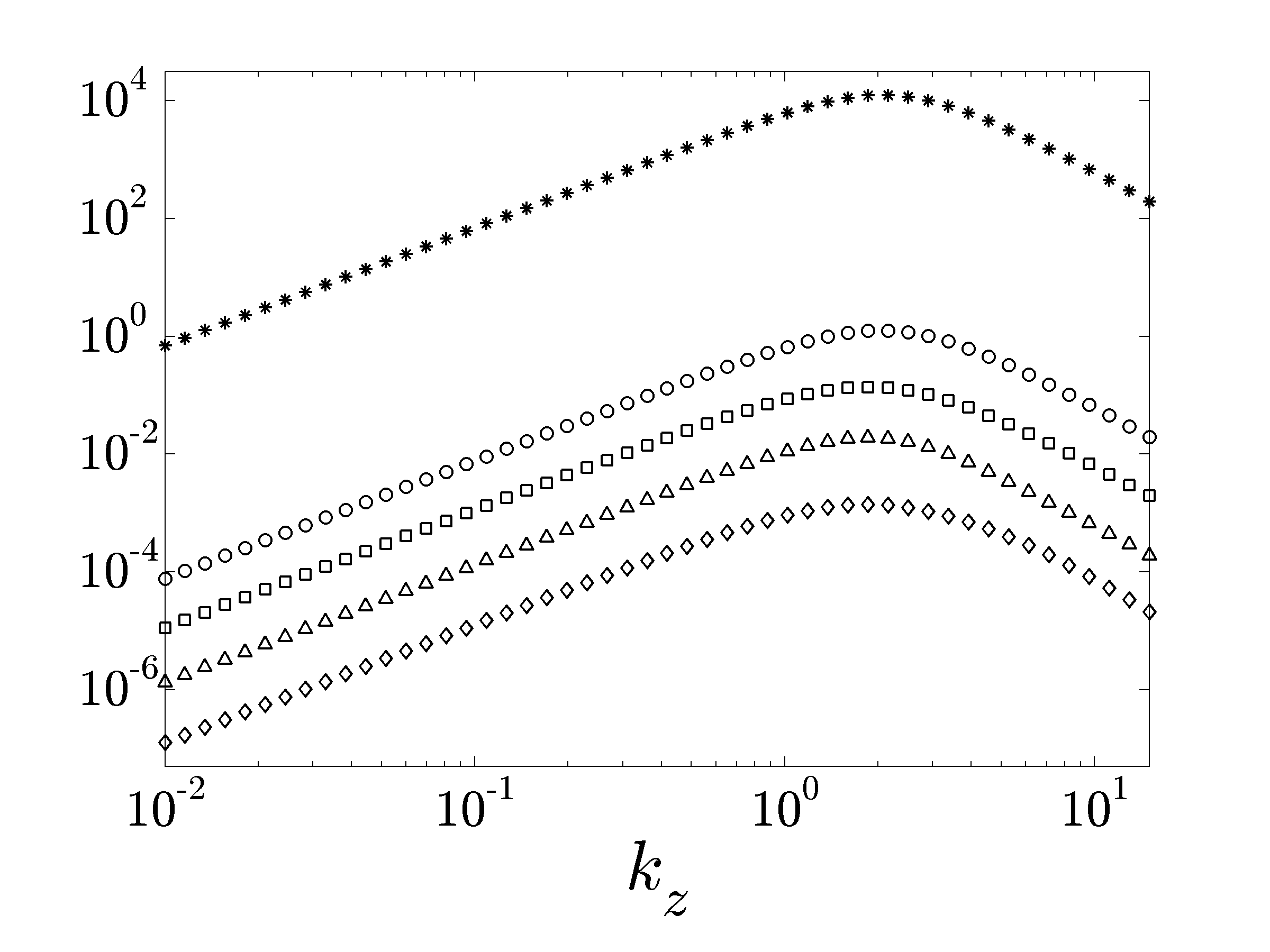}
    &
    \includegraphics[width=0.35\textwidth]{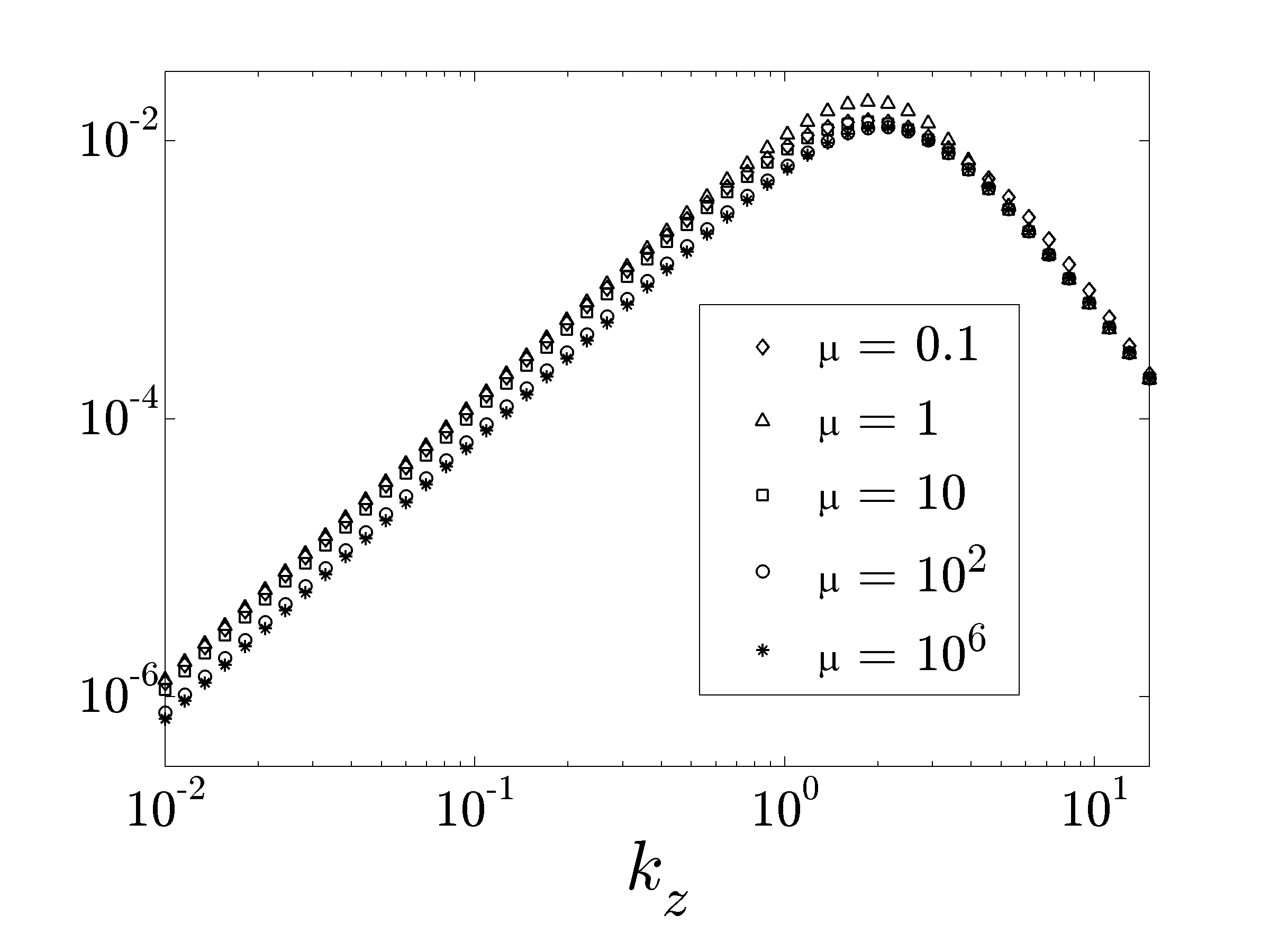}
    \end{tabular}
    \end{center}
    \caption{Variation in $f$, $g$, and $g/\mu$ with $k_z$ for $\beta \, = \, 0.1$ and $\mu \, = \, \{ 0.1$, $1$, $10$, $10^2$, $10^6 \}$; $g$ in Couette
    flow is shown.} \label{Fig:ERemu}
    \end{figure}

We note that function $f$ is the same for all channel flows and only
function $g$ depends on the underlying base flow; for further
analysis of function $g$, we restrict our attention to Couette flow.
Unless noted otherwise, all plots in this section are given in the
log-log scale. Figures~\ref{Fig:ERebeta} and~\ref{Fig:ERemu} show the variations in $f$ and $g$ with $k_z$ at different $\beta$ and $\mu$ values.
Below, we discuss the important observations concerning
the results.

We note that the magnitudes of both $f$ and $g$ increase with an increase
in $\mu$ and a decrease in $\beta$.  This indicates that energy
amplification becomes weaker as one approaches the Newtonian fluid
limit. Furthermore, in all the cases, $f$ monotonically decreases with $k_z$,
while $g$ achieves a maximum at $O(1)$ values of $k_z$. This
suggests that the contribution from $g$ is responsible for the
energy density peaks observed at higher Reynolds
numbers~\cite[][]{Nazish2008a}, confirming our earlier claim that at
higher $Re$, $E \, \approx \, g Re^3$.

Plots in figure~\ref{Fig:ERemu} suggest that $g$ increases
monotonically with $\mu$, while $f$ reaches a saturation limit for
sufficiently large values of $\mu$.  Thus, even in low inertial
regimes, the contribution of $g$ to the energy density can be
significant if the elasticity number, $\mu$, is large enough. In
particular, this demonstrates that energy density peaks observed
by~\citet{Nazish2008a} at $\{ k_x = 0$, $Re = 0.1$, $\beta = 0.1$,
$\mu = 10^6 \}$ arise due to the contribution of $g$.

Moreover, function $g$ approximately scales linearly with $\mu$ for large values of elasticity number.  Figure~\ref{Fig:ERemu} shows the $k_z$-dependence of $g(k_z;\beta,\mu)/\mu$ in Couette flow with $\beta = 0.1$ for five different values of the elasticity number, $\mu \, = \, \{ 0.1$, $1$, $10$, $10^2$, $10^6 \}$. It is evident that the five curves almost collapse onto each other. This is a remarkable discovery in view of rather complicated dependence of the underlying equations on $\mu$ and the range of elasticity numbers considered. Our ongoing theoretical effort is directed towards development of an explicit scaling of $f$ and $g$ with $\mu$; we conjecture that -- for large enough values of elasticity number -- $f$ approximately becomes $\mu$-independent, while $g$ approximately scales linearly with $\mu$. This would suggest the following approximate scaling of the energy density with elasticity number for $\mu \, \gg \, 1$, $E(k_z;Re,\beta,\mu) \, \approx \, Re \tilde{f} (k_z;\beta) \, + \, \mu Re^3 \tilde{g} (k_z;\beta)$, where $\tilde{f}$ and $\tilde{g}$ are $\mu$-independent functions. If this scaling turns out to be correct, our results would indicate an interesting interplay between inertial and viscoelastic effects in energy amplification of Oldroyd-B fluids with low Reynolds/high elasticity numbers~\citep{Larson2000,Groisman2000}.

We have also closely examined the terms in the steady-state Reynolds-Orr
equation that contribute to the $Re^3$-scaling of energy
(results not shown; for details see~\citet{Nazish}).
We observed that the Reynolds-stress and the energy-exchange terms,
$g_{\inprod{u}{U' v}}$ and $g_{\inprod{\tau_{xy}}{\py u}}$, are
positive, suggesting that they lead to energy amplification; the
viscous dissipation and the energy-exchange terms,
$g_{\inprod{u}{\pyy u}}$ and $g_{\inprod{\tau_{xz}}{\mri k_z u}}$,
are negative, suggesting that they lead to energy suppression. Furthermore, the absolute values of all the terms contributing to $g$ increase with an increase in $\mu$ and a decrease in $\beta$. The Reynolds stress term, $g_{\inprod{u}{U' v}}$, appears to reach a saturation limit for sufficiently large $\mu$, while there does not appear to be any upper bound on the absolute values of the other
terms contributing to $Re^3$-scaling as $\mu$ increases. In view of
this, we conclude that the energy-exchange terms are mainly
responsible for energy amplification in flows with pronounced
elasticity effects.

\begin{figure}
    \begin{center}
    \begin{tabular}{cc}
    \includegraphics[width=0.5\textwidth]{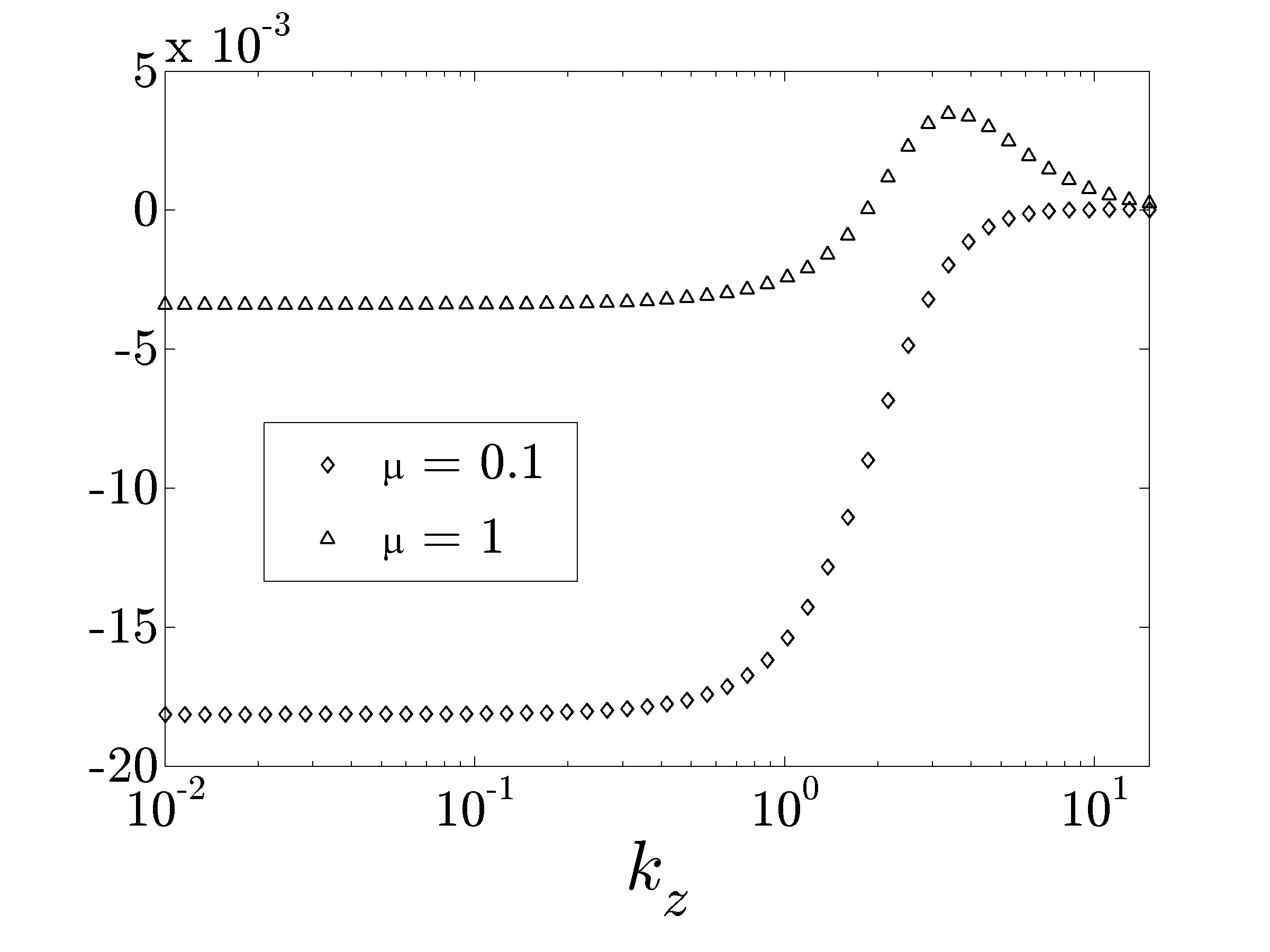}
    &
    \includegraphics[width=0.5\textwidth]{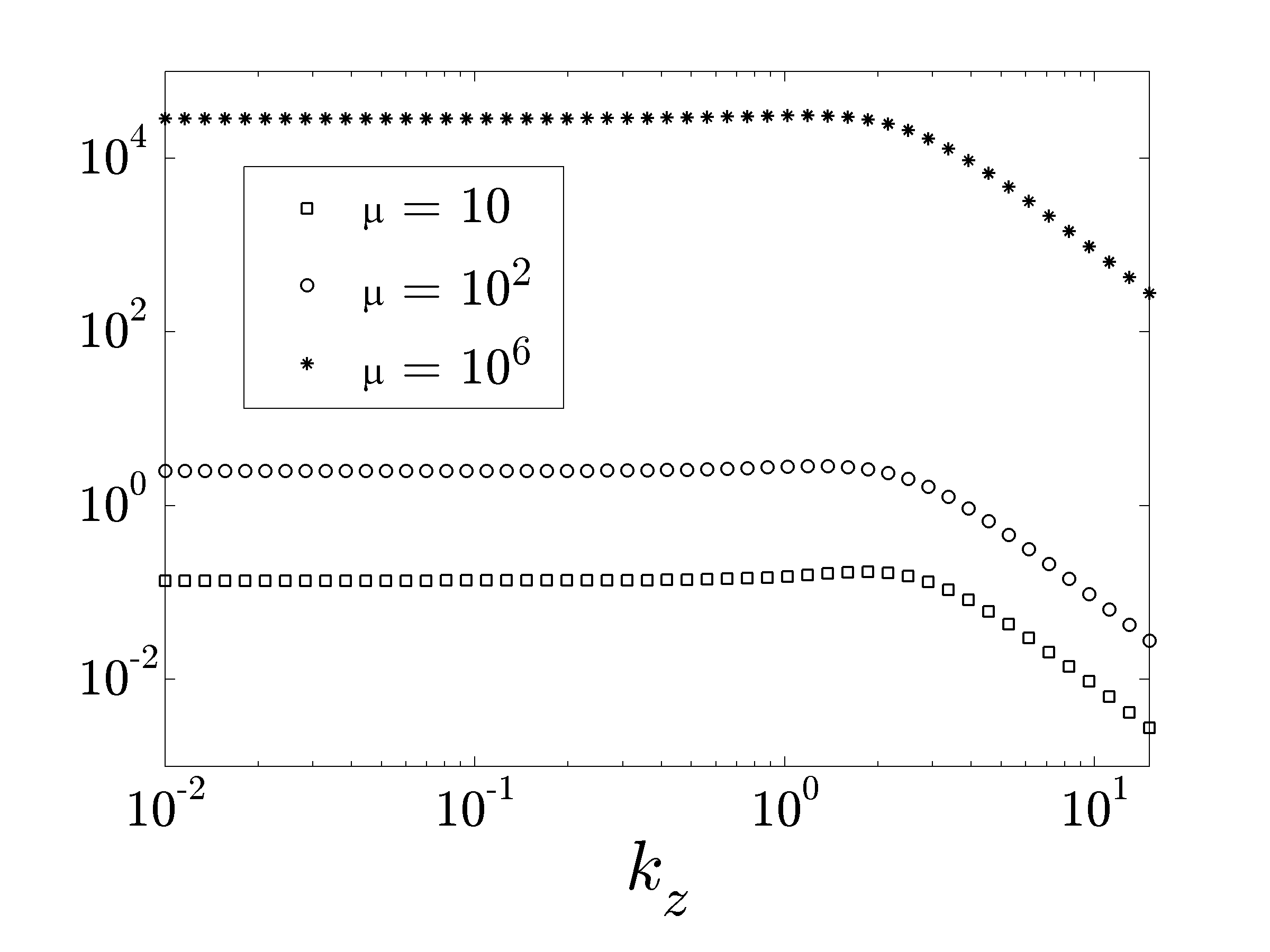}
    \end{tabular}
    \end{center}
    \caption{The $k_z$-dependence of the energy-exchange term
    $g_{\inprod{\tau_{xy}}{\py u}}
    \, + \,
    g_{\inprod{\tau_{xz}}{\mri k_z u}}$ in
    Couette flow with $\beta=0.1$,
    $\mu = \{ 0.1$, $1$, $10$, $10^2$, $10^6 \}$. The left plot is
    shown in the log-linear scale and the right plot is shown in the
    log-log scale.}
    \label{Fig:ERe3energy-exchange}
    \end{figure}

Figure~\ref{Fig:ERe3energy-exchange} shows the $k_z$-dependence of
the energy-exchange term $g_{\inprod{\tau_{xy}}{\py u}}$ $+$
$g_{\inprod{\tau_{xz}}{\mri k_z u}}$ in Couette flow with
$\beta=0.1$ and $\mu = \{ 0.1$, $1$, $10$, $10^2$, $10^6 \}$. This
term contributes to the $Re^3$-scaling of the steady-state energy
and it can lead to energy suppression (for small $\mu$) or energy
amplification (for large $\mu$). Also, the left plot illustrates the
sign-indefiniteness of the energy-exchange term for moderate values
of $\mu$. Clearly, the role of the energy-exchange term becomes more
prominent with increase in elasticity number; in particular, for
$\mu \gg 1$ this term creates a much larger contribution to the
energy amplification than the Reynolds stress term.

It is interesting to contrast the above results with those
of~\citet{Sada2002}, who studied budgets of the
perturbation vorticity and kinetic energy in plane
Poiseuille flow of Oldroyd-B fluids. Their analysis made
use of the linearized equations and a normal-mode decomposition, and considered two-dimensional perturbations. They observed that the temporal frequency
near the critical conditions is a non-monotonic function
of viscosity ratio, and that perturbation shear stresses
have a destabilizing effect.  In our work,
which does not use a normal-mode decomposition and
considers three-dimensional streamwise constant perturbations,
$\Omega_{\max}$ is observed to be a monotonic function
of $\beta$ (but a non-monotonic function of $\mu$), and
perturbation shear stresses are found to create larger energy amplification.
We ascribe the differences in these observations to the different assumptions used in the two studies.

Finally, we briefly comment on the terms that contribute to the
$Re$-scaling of energy. Apart from work done by the body forces and
$-(1-\beta)/(\beta k_z^2) \inprod{\tau_{yz}}{\py w}$, all other
terms contributing to $f$ are negative, which suggests that they
suppress energy amplification; the most dominant of these terms
are $\inprod{w}{\pyy w}$, $\inprod{u_1}{\pyy u_1}$, and
$\inprod{\tau_{xy}}{\py u_1}$. The absolute values of all the terms,
except for the work done by the body forces -- which is independent
of both $\beta$ and $\mu$ -- decrease with an increase in $\beta$.
However, the dependence on $\mu$ is more subtle. The absolute values
of the viscous-dissipation terms contributing to $f$ increase with
an increase in $\mu$, whereas the absolute values of all other terms
decrease. The most striking observation is that the absolute values
of both $\inprod{\tau_{xy}}{\py u_1}$ and $\inprod{\tau_{xz}}{\mri
k_z u_1}$ decrease with an increase in $\mu$, whereas the absolute
values of $\inprod{\tau_{xy}}{\py u_{2,3}}$ and
$\inprod{\tau_{xz}}{\mri k_z u_{2,3}}$ increase with an increase in
$\mu$, suggesting competing effects of these energy-exchange terms
at $Re=1$.

\section{Conclusions}
    \label{sec.conc}

We have investigated the frequency responses of channel flows of
Oldroyd-B fluids.  Our analysis is based on the 2D/3C model, which
simply means that only streamwise-constant perturbations around base
values are considered. The disturbances enter the linearized
governing equations as a body force that can vary in space and time.
The frequency responses are described by an operator having nine
blocks, with a given block relating the component $r$ of the
velocity perturbation to the disturbance in direction $j$.
Characterization of the frequency responses can be performed through
calculation of the Hilbert-Schmidt and $H_2$ norms;
physical interpretations of each of these were discussed.

Our analysis shows that the frequency responses for channel flows of
Oldroyd-B fluids scale with the Reynolds number in exactly the same
way as in Newtonian fluids. Examination of the associated block
diagram reveals that the frequency responses of the spanwise and
wall-normal disturbances to streamwise velocity scale as $Re^2$,
whereas all other frequency responses scale linearly with $Re$.  It
is also seen that streamwise disturbances do not affect the
wall-normal and spanwise velocities. These results indicate that at
high Reynolds numbers, channel flows of both Newtonian and
viscoelastic fluids are most sensitive to spanwise and wall-normal
disturbances, and these disturbances have the largest effect on the
streamwise velocity.  This is also reflected in the behavior of the
various norms, which have the same scalings with $Re$ as is seen for
Newtonian fluids.  The presence of viscoelasticity causes the
frequency responses and their norms to depend on two additional
parameters: the elasticity number and viscosity ratio.

A parametric study of the Hilbert-Schmidt norms, or power spectral
densities, shows that these peak at non-zero temporal frequencies
for viscoelastic fluids. For Newtonian fluids, the peaks occur at
zero frequency, indicating that viscoelasticity reduces the time
scales over which disturbances develop. The frequencies at which the
peaks occur decrease as the solvent contribution to the total
viscosity increases, and the peaks are only present if the
elasticity number is greater than a critical value. An analytical
expression which was derived describes both of these trends
accurately. The peak values of the power spectral densities and the
corresponding spanwise wavenumbers increase with an increase in
elasticity number and a decrease in viscosity ratio. This indicates
that elasticity helps amplify spanwise-dependent disturbances and reduces
their length scale.

Since each component of the frequency response has its own peak in
the Hilbert--Schmidt and $H_2$ norms, different amplification
mechanisms may occur in a given flow. The componentwise analysis
used in the present work makes it easy to detect this, whereas the
aggregate analysis used in our previous work does
not~\cite[][]{Nazish2008a}. The study of the power spectral and
steady-state energy densities also shows that at low Reynolds and
elasticity numbers, the flow is most sensitive to streamwise
disturbances, and these affect the streamwise velocity the most. At
low Reynolds and high elasticity numbers, the flow is most sensitive
to wall-normal and spanwise disturbances, and these again affect the
streamwise velocity the most.

To elucidate how elasticity amplifies disturbances, we have analyzed
a Reynolds-Orr equation for the kinetic energy. Its steady-state
value is exactly the same as the $H_2$ norm. Our analysis shows that
the energy-exchange term involving the polymer stress component
$\tau_{xy}$ and the wall-normal gradient of the streamwise velocity
$\partial_y u$ promotes energy amplification.  In contrast, the
energy-exchange term involving the polymer stress component
$\tau_{xz}$ and the spanwise gradient of the streamwise velocity
$\mri k_z u$ leads to energy suppression. The energy-exchange term
that promotes energy amplification becomes increasingly important
relative to the Reynolds stress term as the elasticity number
increases, and is thus the main driving force for amplification in
flows with strong viscoelastic effects.

The results of the present work significantly extend our knowledge
of how viscoelastic channel flows respond to external disturbances.
They clearly demonstrate how streamwise-constant disturbances can
produce significant energy amplification, and clarify the
relationship between the components of the disturbances and velocity
field.  Notably, the results cover both inertia- and
elasticity-dominated flows. We expect that they will provide a
useful basis for studies exploring nonlinear aspects of transition
to inertial and/or elastic turbulence in channel flows of
viscoelastic fluids.  It might also be possible to extend
the ideas developed here to study interfacial instabilities,
where the linearized governing equations are also non-normal
in nature~\cite*[][]{Lin2004}.

    \begin{acknowledgments}
The work of N.\ H.\ and S.\ K.\ was partially supported by the
Donors of The American Chemical Society Petroleum Research Fund. The
work of M.\ R.\ J.\ was partially supported by the National Science
Foundation under CAREER Award CMMI-06-44793.
    \end{acknowledgments}

%%============
%% Appendix %%
%%============

\appendix

\section{Frequency response operators}
    \label{append.FR}

The Reynolds-number-independent frequency response operators
$\bar{\bH}_{rj}$ can be obtained by applying the temporal Fourier
transform to Eqs.~(\ref{eq.lnse-2d3c}).
Equation~(\ref{eq.lnse-2d3c-tauv}) can be used to express polymer
stresses
    $
    \bphi_2
    \, = \,
    \left[\,\tau_{yy}~\tau_{yz}~\tau_{zz}\,\right]^T
    $
in terms of wall-normal velocity
    $
    \phi_1
    \, = \,
    v
    $
    \beq
    \bphi_2
    \; = \;
    \dfrac{\bF_{21}}{1 \, + \, \mri \mu \Omega} \phi_1,
    ~~~
    \Omega
    \; = \;
    \omega Re.
    \label{eq.phi2}
    \eeq
Substitution of Eq.~(\ref{eq.phi2}) into the Fourier transform
of Eq.~(\ref{eq.lnse-2d3c-v}) yields
    \beq
    \phi_1
    \; = \;
    Re
    \left(
    \mri \Omega \bI
    \, - \,
    \beta \bF_{11}
    \, - \,
    \dfrac{1 \, - \, \beta}{1 \, + \, \mri \mu \Omega}
    \bF_{12} \bF_{21}
    \right)^{-1}
    \left(
    \bB_2 d_2 \, + \,
    \bB_3 d_3
    \right).
    \label{eq.phi1}
    \eeq
Using Eq.~(\ref{eq.lnse-2d3c-out}) and the fact that
    $
    \bF_{12} \bF_{21}
    \, = \,
    \bF_{11}
    \, = \,
    \bDelta^{-1} \bDelta^2
    $,
the expressions for operators
    $
    \bar{\bH}_{r j}(k_z,\Omega;\beta,\mu),
    $
    $
    \{r = v,w; \; j = 2,3 \},
    $
are readily established (cf.\ \S~\ref{sec.frRe}).

The following relation between the wall-normal velocity/vorticity
($\phi_1,\phi_3$) and polymer stresses
    $
    \bphi_4
    \, = \,
    \left[\,\tau_{xy}~\tau_{xz}\,\right]^T
    $
can be established by substituting Eq.~(\ref{eq.phi2}) to the Fourier
transform of Eq.~(\ref{eq.lnse-2d3c-taueta})
    \beq
    \bphi_4
    \; = \;
    \dfrac{\mu Re}{1 \, + \, \mri \mu \Omega}
    \left(
    \bF_{41}
    \, + \,
    \dfrac{\bF_{42} \bF_{21}}{1 \, + \, \mri \mu \Omega}
    \right)
    \phi_1
    \; + \;
    \dfrac{1}{1 \, + \, \mri \mu \Omega}
    \bF_{43} \phi_3.
    \non
    \eeq
Substituting the above equation in the Fourier transform
of Eq.~(\ref{eq.lnse-2d3c-eta}) yields
    \beq
    \ba{rcl}
    \phi_3
    & \!\! = \!\! &
    Re
    \left(
    \mri \Omega \bI
    -
    \beta \bF_{33}
    -
    \dfrac{(1 - \beta)}{1 + \mri \mu \Omega} \bF_{34} \bF_{43}
    \right)^{-1}
    \left(
    \bF_{31}
    +
    \dfrac{\mu (1 - \beta) \bF_{34}}{1 + \mri \mu \Omega}
    \left(
    \bF_{41} + \dfrac{\bF_{42} \bF_{21} }{1 + \mri \mu \Omega}
    \right)
    \right) \phi_1
    \\[0.25cm]
    & \!\! + \!\! &
    Re
    \left(
    \mri \Omega \bI
    -
    \beta \bF_{33}
    -
    \dfrac{(1 - \beta)}{1 + \mri \mu \Omega} \bF_{34} \bF_{43}
    \right)^{-1}
    \bB_1 d_1.
    \ea
    \label{eq.phi3}
    \eeq
Now, since $u \, = \, \bC_u \phi_3$, the expressions for operators
$\bar{\bH}_{u j}(k_z,\Omega;\beta,\mu)$, $\{ j = 1,2,3 \}$, given in
\S~\ref{sec.frRe} follow immediately from this equation and the fact
that in Couette and Poiseuille flows we have
    $
    \{
    {\bF}_{34} {\bF}_{43}
    =
    {\bF}_{33}
    =
    \bDelta,
    $
    $
    {\bF}_{34} {\bF}_{41}
    =
    0,
    $
    $
    {\bF}_{34} {\bF}_{42} {\bF}_{21}
    =
    \mri k_z
    \left( U'(y) \bDelta
    +
    2 U''(y) \py
    \right)
    \}.
    $
The latter expression appears in the coupling operator
$\bC_{p2}$ of \S~\ref{sec.frRe}. A careful examination
of the governing equations shows that this term arises
due to the tensor involving polymer stress fluctuations and
gradients in the base velocity profile (the $\btau \cdot \mbox{\boldmath$\nabla$} \mathbf{\overline{v}}$ term in Eq.~(\ref{eq.linear-original})).

\section{Determination of the $H_2$ norm using Lyapunov equation}
    \label{sec.lyap}

Next, we outline the procedure that is most convenient for determination of the
Reynolds-number-independent functions $f$ and $g$ in the expression for the energy amplification. This method exploits the fact that the
steady-state energy density of the frequency response operator~$\bH$
can be determined from the Lyapunov equation, and it is very
suitable for uncovering the explicit scaling with $Re$ of various
terms in the steady-state Reynolds-Orr equation.

\subsection{Lyapunov equation and steady-state energy density}

Equations~(\ref{eq.lnse-2d3c-v})-(\ref{eq.lnse-2d3c-taueta}) from
\S~\ref{sec.dff} can be compactly rewritten as
    \beq
    \tbo{\pt \, \bvphi_1}{\pt \, \bvphi_2}
    \; = \;
    \tbt{(1/Re) \bA_{11}}{0}{\bA_{21}}{(1/Re) \bA_{22}}
    \tbo{\bvphi_1}{\bvphi_2}
    \; + \;
    \tbo{\bB_{11}}{\bB_{21}}
    \bd,
    \non
    \eeq
where $\bvphi_1 = \left[\,\phi_1~~\bphi_2^T\,\right]^T$, $\bvphi_2 =
\left[\,\phi_3~~\bphi_4^T\,\right]^T$, and
    \beq
    \ba{c}
    \bA_{11}
    \; = \;
    \tbt{\beta \bF_{11}}{(1 - \beta) \bF_{12}}{(1/\mu) \bF_{21}}{-(1/\mu) \bI},
    ~~
    \bA_{22}
    \; = \;
    \tbt{\beta \bF_{33}}{(1 - \beta) \bF_{34}}{(1/\mu) \bF_{43}}{-(1/\mu) \bI},
    \\[0.35cm]
    \bA_{21}
    \; = \;
    \tbt{\bF_{31}}{0}{\bF_{41}}{\bF_{42}},
    ~~
    \bB_{11}
    \; = \;
    \tbth{0}{\bB_2}{\bB_3}{0}{0}{0},
    ~~
    \bB_{21}
    \; = \;
    \tbth{\bB_1}{0}{0}{0}{0}{0}.
    \ea
    \non
    \eeq
Furthermore, the structure of operators $\bB_{11}$ and $\bB_{21}$
yields
    \beq
    \bM
    \;=\;
    \tbo{\bB_{11}}{\bB_{21}}
    \obt{\bB_{11}^*}{\bB_{21}^*}
    \; = \;
    \tbt{\bM_{11}}{0}{0}{\bM_{22}},
    \non
    \eeq
where the asterisk denotes the adjoint of a given operator. From the
definitions of operators $\bB_j$, $j = 1,2,3$, and their respective
adjoints~\cite[][]{Mihailo2005} we have
    \beq
    \bM_{11}
    \,=\,
    \tbt{\bB_2 \bB_2^* + \bB_3 \bB_3^*}{0}{0}{0}
    \,=\,
    \tbt{\bI}{0}{0}{0},
    ~~
    \bM_{22}
    \,=\,
    \tbt{\bB_1 \bB_1^*}{0}{0}{0}
    \,=\,
    \tbt{\bI}{0}{0}{0}.
    \non
    \eeq
The partition of operators $\bM_{11}$ and $\bM_{22}$ is done
conformably with the partition of vectors $\bvphi_1 =
\left[\,\phi_1~~\bphi_2^T\,\right]^T$ and $\bvphi_2 =
\left[\,\phi_3~~\bphi_4^T\,\right]^T$, respectively.

The steady-state energy density (that is, the $H_2$ norm of operator
$\bH$) can be expressed in terms of the solution to the following
operator Lyapunov equation
    \beq
    \bA  \bP
    \;+\;
    \bP
    \bA^*
    \;=\;
    - \bM
    \non
    \eeq
as
    $
    E(k_z;Re,\beta,\mu)
    \, = \,
    \trace \, ( \bN \bP ),
    $
where $\bP$ and $\bN$ are $2 \times 2$ self-adjoint block operators
with the following structure
    \beq
    \ba{c}
    \bP
    \; = \;
    \tbt{\bP_{11}}{\bP_{21}^*}{\bP_{21}}{\bP_{22}},
    ~~
    \bN
    \; = \;
    \tbt{\bN_{11}}{0}{0}{\bN_{22}},
    \\[0.35cm]
    \bN_{11}
    \,=\,
    \tbt{\bC_v^* \bC_v + \bC_w^* \bC_w}{0}{0}{0}
    \,=\,
    \tbt{\bI}{0}{0}{0},
    ~~
    \bN_{22}
    \,=\,
    \tbt{\bC_u^* \bC_u}{0}{0}{0}
    \,=\,
    \tbt{\bI}{0}{0}{0}.
    \ea
    \non
    \eeq
Using the structure of operators $\bP$ and $\bN$, it follows that
$E(k_z;Re,\beta,\mu)$ is determined by
    $
    E (k_z;Re,\beta,\mu)
    \, = \,
    \trace \, ( \bN \bP)
    \, = \,
    \trace \, ( \bN_{11} \bP_{11})
    +
    \trace \, ( \bN_{22} \bP_{22}).
    $
The same argument as in~\citet[][]{Bamieh2001,Mihailo2005} can be used to
show
    $
    \{
    \bP_{11}
    \, = \,
    Re \, \bX,
    $
    $
    \bP_{21}
    \, = \,
    Re^2 \, \bY,
    $
    $
    \bP_{22}
    \, = \,
    Re
    \,
    \bZ_1
    \, + \,
    Re^3
    \,
    \bZ_2
    \},
    $
where the Reynolds-number-independent operators $\bX$, $\bY$, and
$\bZ$ satisfy the following system of conveniently coupled equations
    \begin{subequations}
    \begin{align}
    \bA_{11}  \bX
    \; + \;
    \bX  \bA_{11}^*
    & \;=\;
    -
    \bM_{11},
    \non
    \\
    \bA_{22}  \bY
    \; + \;
    \bY  \bA_{11}^*
    & \;=\;
    -  \bA_{21}  \bX,
    \non
    \\
    \bA_{22}  \bZ_2
    \; + \;
    \bZ_2  \bA_{22}^*
    & \;=\;
    - (  \bA_{21}  \bY^* \; + \;  \bY  \bA_{21}^*),
    \non
    \\
    \bA_{22}  \bZ_1
    \; + \;
    \bZ_1  \bA_{22}^*
    & \;=\;
    -  \bM_{22}.
    \non
    \end{align}
    \non
    \end{subequations}
Hence,
    \beq
    \ba{rcl}
    E (k_z;Re,\beta,\mu)
    & = &
    Re
    \left(
    \trace \, ( \bN_{11} \bX )
    \, + \,
    \trace \, ( \bN_{22} \bZ_1)
    \right)
    \; + \;
    Re^3 \, \trace \, ( \bN_{22} \bZ_2)
    \\
    & = &
    Re \, f(k_z;\beta,\mu)
    \; + \;
    Re^3 \, g(k_z;\beta,\mu),
    \ea
    \non
    \eeq
which provides an efficient way of computing functions $f$ and $g$ in the expression for the steady-state energy density.

\subsection{Lyapunov equation and steady-state Reynolds-Orr equation}

We next illustrate how the terms on the right-hand side of the
steady-state Reynolds-Orr equation can be obtained from the solution
of the Lyapunov equation. This allows us to determine the explicit
$Re$-scaling of these terms, which clarifies the importance of different
energy amplification mechanisms.

We begin by observing that the solution of Lyapunov equation, $\bP$,
represents the steady-state correlation operator of $\bvphi =
\left[\,\bvphi_1^T~~\bvphi_2^T\,\right]^T$, that is
    \[
    \bP
    \; = \;
    \lim_{t \, \rightarrow \, \infty}
    {\cal E}
    \left\{
    \bvphi (\cdot,t)
    \otimes
    \bvphi (\cdot,t) \right\},
    \]
where $\bvphi \otimes \bvphi$ denotes the tensor product of $\bvphi$
with itself. This operator contains all the second-order
steady-state statistics of the velocity and polymer stress fields.
For example, operator $\bP_{22}$ carries information about
the steady-state correlation of $\bvphi =
\left[\,\omega_y~~\tau_{xy}~~\tau_{xz}\,\right]^T$ with itself, and
a simple kinematic relationship between $\omega_y$ and $u$ (at $k_x
= 0$, $u = \bC_u \omega_y = -(\mri/k_z) \omega_y$) can be used to
obtain all possible correlations between $u$, $\tau_{xy}$, and
$\tau_{xz}$. In particular, owing to the convenient scaling of
$\bP_{22}$ with the Reynolds number
    ($
    \bP_{22}
    =
    Re
    \,
    \bZ_1
    \, + \,
    Re^3
    \,
    \bZ_2
    $),
we conclude that all these correlations contain a part that scales
as $Re$ and a part that scales as $Re^3$. (The streamwise force is
responsible for the $Re$-scaling and the wall-normal and spanwise forces
are responsible for the $Re^3$-scaling.) Now, since the inner product of
two fields is equal to the trace of their outer (tensor) product, we
conclude that the terms $\inprod{u}{\pyy u}$,
$\inprod{\tau_{xy}}{\py u}$, and $\inprod{\tau_{xz}}{\mri k_z u}$ in
the steady-state Reynolds-Orr equation scale as $f_i (k_z;\beta,\mu)
Re + g_i (k_z;\beta,\mu) Re^3$, $\{i = \inprod{u}{\pyy u}$,
$\inprod{\tau_{xy}}{\py u}$, $\inprod{\tau_{xz}}{\mri k_z u} \}$.
Since the Reynolds stress term $-\inprod{u}{U'v}$ can be determined
from $\bP_{21} = Re^2 \bY$, it follows that its contribution to the
steady-state energy scales as $Re^3$ (cf.\ \S~\ref{sec.Energy}). The
contributions of all other terms scale as $Re$.

\end{document}